\documentclass[twocolumn,twocolappendix,trackchanges]{aastex63}
\usepackage{bm}
\usepackage{braket}
\usepackage{amsmath}
\usepackage{subeqnarray}
\usepackage{here}
\usepackage{lineno}

\received{January 29, 2022}
\revised{June 27, 2022}
\accepted{July 3, 2022}
\submitjournal{ApJ}
\shorttitle{}
\shortauthors{Shimada et al.}
\begin{document}
\title{MEAN-FIELD ANALYSIS ON LARGE-SCALE MAGNETIC FIELDS AT HIGH REYNOLDS NUMBERS}

\correspondingauthor{Ryota Shimada}
\email{shimada\_ryota@eps.s.u-tokyo.ac.jp}

\author[0000-0002-8507-3633]{Ryota Shimada}
\affiliation{\rm{Department of Earth and Planetary Science, The University
of Tokyo, 7-3-1, Hongo, Bunkyo-ku, Tokyo, 113-0033, Japan.}}

\author[0000-0002-6312-7944]{Hideyuki Hotta}
\affiliation{\rm{Department of Physics, Graduate School of Science, Chiba
University, 1-33 Yayoi-cho, Inage-ku, Chiba 263-8522, Japan}}

\author[0000-0001-5457-4999]{Takaaki Yokoyama}
\affiliation{\rm{Astronomical Observatory, Kyoto University, Sakyo-ku, Kyoto, 606-8502, Japan}}

\begin{abstract}
Solar magnetic fields comprise an 11-year activity cycle, represented by the number of sunspots. The maintenance of such a solar magnetic field can be attributed to fluid motion in the convection zone, i.e. a dynamo. This study conducts the mean-field analyses of the global solar dynamo simulation presented by \citet{2016Sci...351...1427}.
Although the study succeeds in producing coherent large-scale magnetic fields at high Reynolds numbers, the detailed physics of the maintenance of this field have not been fully understood.
This study extracts the $\alpha$-tensor and the turbulent magnetic diffusivity tensor $\bm{\beta}$ through mean-field analyses.
The turbulent magnetic diffusivity exhibits a significant decrease towards high Reynolds numbers.
The decrease in the turbulent magnetic diffusivity suppresses the energy conversion of large-scale field to small-scale field.
This implies that the decrease in the turbulent magnetic diffusivity contributes to the maintenance of a large-scale magnetic field at high Reynolds numbers. 
A significant downward turbulent pumping is observed; it is enhanced in the weak phase of the large-scale field. 
This study proposes a cyclic reversal process of a large-scale field which is dominantly driven by the $\alpha$-effect and is possibly triggered by downward pumping.
\end{abstract}


\section{Introduction\label{sec:intro}} 
Solar magnetic fields comprise an 11-year activity cycle, represented by the number of sunspots.
They are also characterized by their large-scale coherent structures and polarity reversals.
The prominent examples are the 11-year polar field reversals and the sunspot parity rules \citep{1919_apj_49_153}.
The maintenance of such a solar magnetic field can be attributed to fluid motion in the convection zone, i.e. a dynamo.

\begin{deluxetable*}{cccc}
  \tablecaption{Lists of calculated cases.\label{tab: cases}}
  \tablecolumns{4}
  \tablenum{1}
  \tablewidth{0pt}
  \tablehead{
  \colhead{Case} &
  \colhead{Low} &
  \colhead{Mid} & 
  \colhead{High} }
  \startdata
  $(N_r,N_\theta,N_\varphi)$ \tablenotemark{$^\mathrm{(1)}$} & $(64\times192\times384)$ & $(64\times192\times384)$& $(192\times384\times768)$ \\
  Diffusivities \tablenotemark{$^\mathrm{(2)}$} [cm$^2$\ s$^{-1}$]& 1$\times10^{12}$ & 0 & 0 \\
  Run times [yr]& 117 & 36 & 31 \\
  Rm \tablenotemark{$^\mathrm{(3)}$}& 319 & 336 & 551 \\
  Pm \tablenotemark{$^\mathrm{(3)}$}& 1 & 1.40 & 1.77 \\
  $E_{\mathrm{turb}}$ \tablenotemark{$^\mathrm{(4)}$} [$10^6$ erg\ cm$^{-3}$]& 1.6 &2.3 & 2.7 \\
  $E_{\mathrm{mean}}$ \tablenotemark{$^\mathrm{(4)}$} [$10^4$ erg\ cm$^{-3}$]& 16.5 & 9.7 & 9.8 \\
  \enddata
  \tablenotetext{}{(1) The number of grid points are counted on the spherical coordinate projected from Yin-Yang grid \citep{2004_ggg_5_9005}. (2) Imposed explicit diffusivities at the top boundary are presented in the tables. The explicit viscosity and explicit diffusivity have identical values. All cases also include the artificial viscosity reported by \citet{Rempel_2014}. (3) The magnetic Reynolds number (Rm) and magnetic Prandtl number (Pm) are estimated from the energy spectra \citep{2016Sci...351...1427}. (4) The energy densities of the small-scale magnetic field ($E_{\mathrm{turb}}$) and large-scale field ($E_{\mathrm{mean}}$) at the base of the convection zone ($0.71R_\sun <r<0.73R_\sun$) are presented. These values are considered as temporal average over 5.5-27.4 yr.}
  \end{deluxetable*}

\citet{1983_apjs_53_243} and \citet{1984_JCohP_55_461,1985_apj_55_461} conducted studies using global three-dimensional (3D) magnetohydrodynamics (MHD) simulation.
Furthermore, various 3D MHD simulations have been conducted in the past decade \citep{2011ApJ...735..46P,2012_ApJ_762_73P,2013_apj_778_11,2014_apj_789_35,2015_apj_789_35,2016_MNRAS_456_1708,2016_apj_819_104,2016Sci...351...1427,2017_aap_599_A4,2018_apj_863_35,2021_natas}.
Although realistic Reynolds numbers are large values of $\rm{Re}\backsimeq10^{13}$ and $\rm{Rm}\backsimeq10^{10}$ at the base of the convection zone \citep{2003...aapr...11...287}, most of the calculations used those of lower values of approximately 100-300 \citep{2020...LRSP...17...1}, i.e. they used large viscosity and magnetic diffusivity ($\backsimeq10^{12}$ cm$^2$ s$^{-1}$) to suppress the small-scale fluctuating field and obtain a coherent large-scale magnetic field.
The main drawback found by previous studies was that the energy and coherence of the large-scale magnetic field were destroyed at higher Reynolds numbers.  
In this context, \citet{2016Sci...351...1427} succeeded in producing a coherent large-scale magnetic field at higher Reynolds numbers using high-resolution calculations.
They reported that an efficient small-scale dynamo suppresses the small-scale flow, which consequently maintains the large-scale magnetic field. 
However, the specific physics involved in maintaining the large-scale magnetic field and reversing its polarity must be further analyzed. 

Other approach to study the solar dynamo is the mean-field electrodynamics \citep{1966...ZN...369...21a}, in which the large-scale magnetic field is considered as the mean field.
This approach is useful in analyzing the physics of the 3D MHD calculations due to its simplicity.
\citet{2011ApJ...735..46P} developed a method which extracts the mean-field parameters, that is, the $\alpha$-tensor from the calculation using singular value decomposition. 
The mean-field approach was used to identify the major factors contributing to the induction of a large-scale magnetic field \citep[see][]{2010_ApJ_711_424P, 2012_ApJ_762_73P, 2015_ApJ_809_149P}. 

This study aims to understand the physics involved in maintaining a coherent large-scale magnetic field at high Reynolds numbers. 
The mean-field analyses of the results of the 3D MHD simulations \citep{2016Sci...351...1427} are conducted for this purpose.   
Particularly, the $\alpha$-tensor and $\beta$-tensor are extracted, and each term of the induction equation of the mean-magnetic field is estimated.  
Section \ref{sec:overview of the simulation} presents the basic settings of the simulation and the realization of a large-scale field.
Section \ref{sec:mean-field analysis} presents the procedures and results of the mean-field analyses. 
Section \ref{sec:discussion and conclusion} presents the maintenance and polarity reversal of a global-scale magnetic field at high Reynolds numbers. 
Lastly, Section \ref{subsec:conclusion} presents the conclusion.

In the course of our analysis and discussion, the radial pumping ($\gamma_r$) grabs the spotlight.
The effect of the radial pumping on the solar cycle is first examined by \citet{Brandenburg_1992} using the mean-field simulation.
The background of this study is the existence of radial pumping in the convection zone suggested by \citet{Nordlund_1992}, which conducts local Cartesian simulation.
\citet{Brandenburg_1992} artificially select the value of $\gamma_r$ in their simulation at that time.
Estimation of the radial pumping in 3D MHD simulation started by \citet{ossendrijver_2002} using local Cartesian simulation.
In our work, the global 3D MHD simulation is used to obtain the global distribution of $\gamma_r$. 
This allows us to estimate the distribution of induction by radial pumping and gain insight into polarity reversal.
Furthermore, it is noteworthy that $\gamma_r$ under the several Reynolds numbers are extracted and discussed in this work.

\section{OVERVIEW OF THE SIMULATION} \label{sec:overview of the simulation}  
This section presents a detailed description of the basic setting of the simulation and the realized large-scale field.

\subsection{Basic settings}
\begin{figure*}
  \epsscale{0.5}
  \gridline{\fig{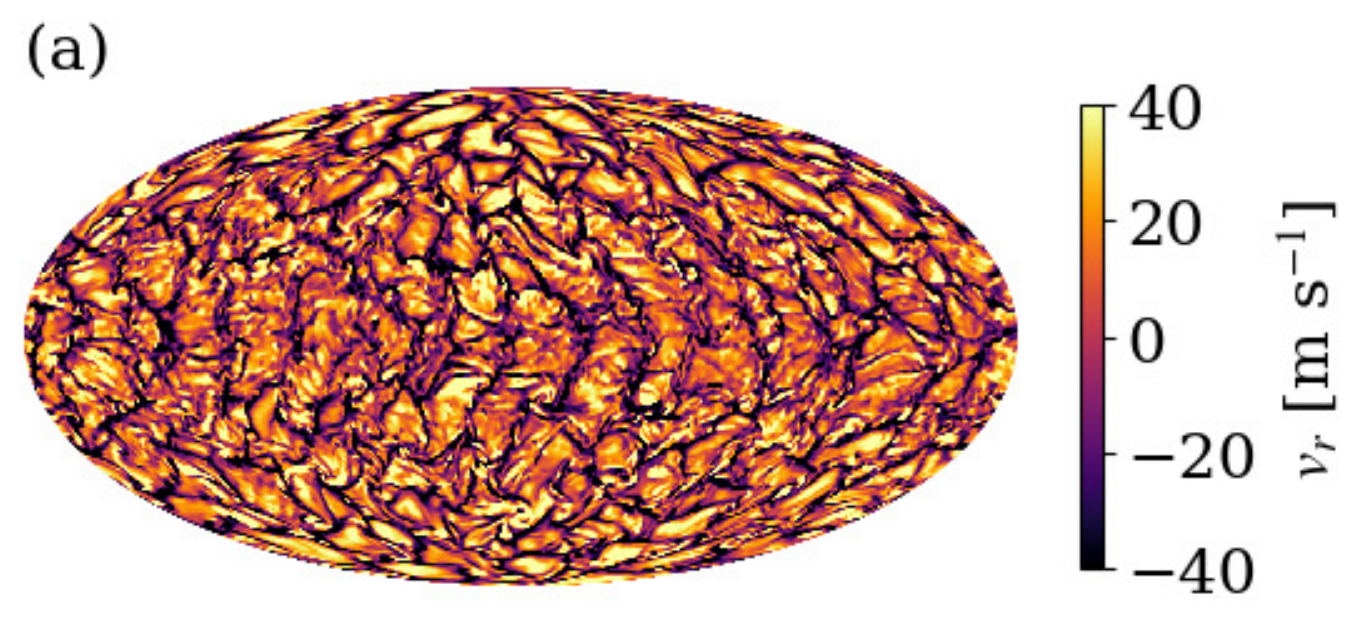}{0.5\textwidth}{}
  \fig{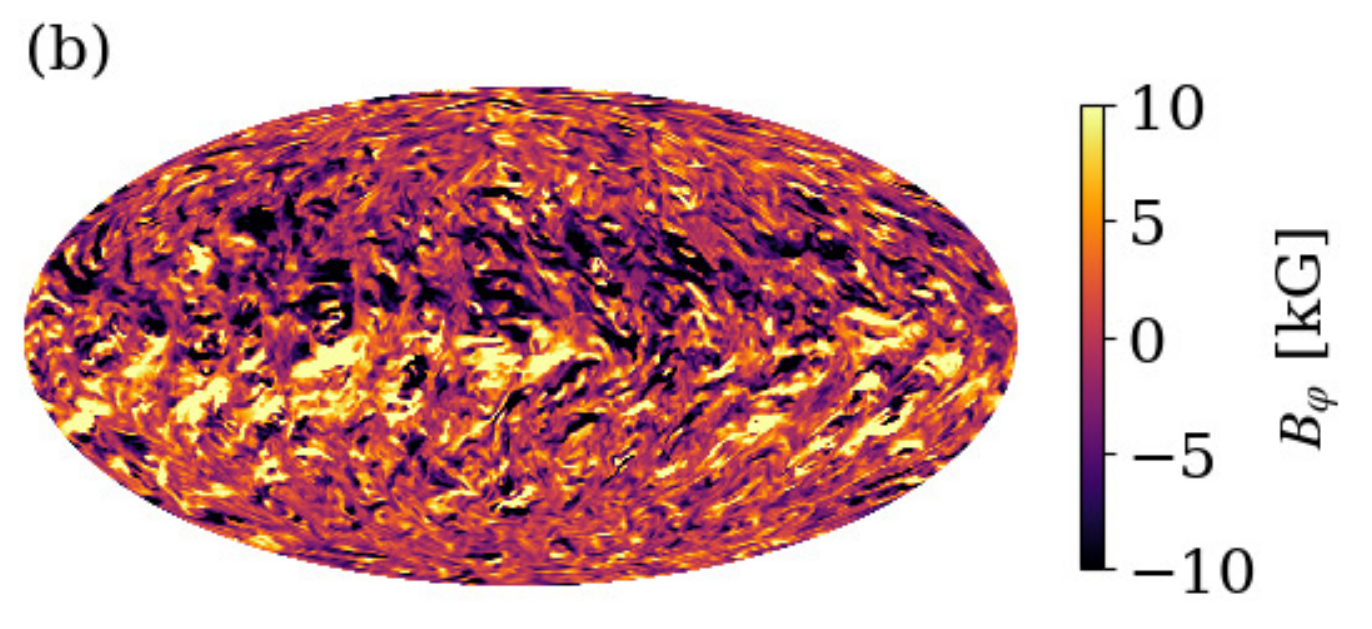}{0.5\textwidth}{}
  }
  \caption{Snapshot of (a) radial velocity $v_r$ and (b) logitudinal magnetic field $B_\varphi$  at $t=5.5$ yr when large-scale field is constructed.}
  \label{fig:snapshot}
\end{figure*}

The data are calculated using a similar approach as in \citet{2016Sci...351...1427}.
The simulation adopts the reduced speed of sound technique (RSST) \citep{2012_aap_539_A30} and solves the 3D MHD equations
in spherical geometry ($r, \theta, \varphi$) with gravity and rotation as follows:
\begin{eqnarray}
  \frac{\partial \rho_1}{\partial t}=&&-\frac{1}{\xi^2}\nabla\cdot(\rho\bm{v})\label{eq:eqs1},\\
  \rho\frac{\partial \bm{v}}{\partial t}=&&-\rho(\bm{v}\cdot\nabla)\bm{v}+2\rho\bm{v}\times\bm{\Omega}_0\nonumber-\nabla\left(p_1+\frac{B^2}{8\pi}\right)\nonumber\\
  &&+\nabla\cdot\left(\frac{\bm{B}\bm{B}}{4\pi}\right)-\rho_1g\bm{e}_r-\nabla\cdot\bm{D}\label{eq:eqs2},\\
  \rho T\frac{\partial s_1}{\partial t}=&&-\rho T(\bm{v}\cdot\nabla)s+\frac{1}{r^2}\frac{d}{dr}\left(r^2\kappa_r\rho_0c_p\frac{dT_0}{dr}\right)\nonumber\\
  &&+\Gamma(r)+\nabla\cdot(\kappa\rho_0T_0\nabla s)\nonumber\\
  &&+2\rho \nu\left[e_{ij}e_{ij}-\frac{1}{3}(\nabla\cdot\bm{v})^2\right]\nonumber\\
  &&+\frac{\eta}{4\pi}(\nabla\times\bm{B})^2\label{eq:eqs3},\\
  \frac{\partial \bm{B}}{\partial t}=&&\nabla\times(\bm{v}\times\bm{B}-\eta\nabla\times\bm{B})\label{eq:eqs4},\\
  p_1=&&\left(\frac{\partial p}{\partial \rho}\right)_s\rho_1+\left(\frac{\partial p}{\partial s}\right)_\rho s_1\label{eq:eqs5},
\end{eqnarray} 
where $\rho$, $p$, $s$, $T$, $\bm{v}$, and $\bm{B}$ represent the density, gas pressure, specific entropy, temperature, fluid velocity, and magnetic field, respectively.
The radial extent is restricted to $r_1 < r < r_2\ (r_1=0.71R_\sun, r_2=0.96R_\sun)$.
The subscript, 0, denotes the background spherically symmetric stratification; the solar standard model \citep[Model S: ][]{1996_sci_5256_1286} is employed for this calculation.
The subscript, 1, denotes the fluctuation from the background stratification. 
$\xi$ is the parameter in the RSST, which reduces the effective speed of sound by a factor of $\xi$.
The gravitational acceleration, $g$, and radiative diffusivity, $\kappa_r$, are adopted from Model S.
The rotation rate is set as the solar rotation rate, $|\bm{\Omega}_0|/(2\pi)=413$ nHz \citep{2003_araa_41_599}. 
$\Gamma(r)$ is the cooling term, which is effective only near the surface.
$\bm{D}$ and $e_{ij}$ represent the viscous stress tensor and strain rate tensor, respectively.
A strong thermal conductivity $\kappa=2\times10^{13}$ cm$^2$\ s$^{-1}$ at the top boundary is adopted to obtain a solar-like differential rotation. 
The thermal conductivity, $\kappa$, explicit viscosity, $\nu$, and explicit magnetic diffusivity, $\eta$, are set to have a radial dependence of $1/\sqrt{\rho_0}$ \citep{2014_apj_789_35}.

The analyzed cases are listed in Table \ref{tab: cases}. These cases are distinguished by the number of grid points and the imposed diffusivities. Case High contains twice as many grid points as the other cases in each direction. 
The explicit diffusivities, i.e., magnetic diffusivity and viscosity, are imposed in case Low, and only numerical diffusivities exist in the other cases.

The estimated magnetic Reynolds number (Rm) and magnetic Prandtl number (Pm) are listed in Table \ref{tab: cases}. 
The magnetic Reynolds and Prandtl numbers are evaluated using the energy spectra in cases Mid and High since no explicit diffusivities are used and the dissipations are all produced by numerical ones \citep[see][]{2016Sci...351...1427}.
The magnetic Reynolds number increases slightly from Low to Mid and significantly increases toward High.

The main differences between this simulation and \citet{2016Sci...351...1427} are the artificial diffusivity and the number of grid points. 
The artificial diffusivity suggested by \citet{Rempel_2014} is implemented on all physical quantities in ours, whereas one suggested by \citet{Rempel_2009} is used exceptionally on density in \citet{2016Sci...351...1427} . 
The number of grid points for each direction in our case High is half as many as the counterpart.

\subsection{Large-scale field}\label{subsec:large-scale field}
\begin{figure}
  \epsscale{1}
  \plotone{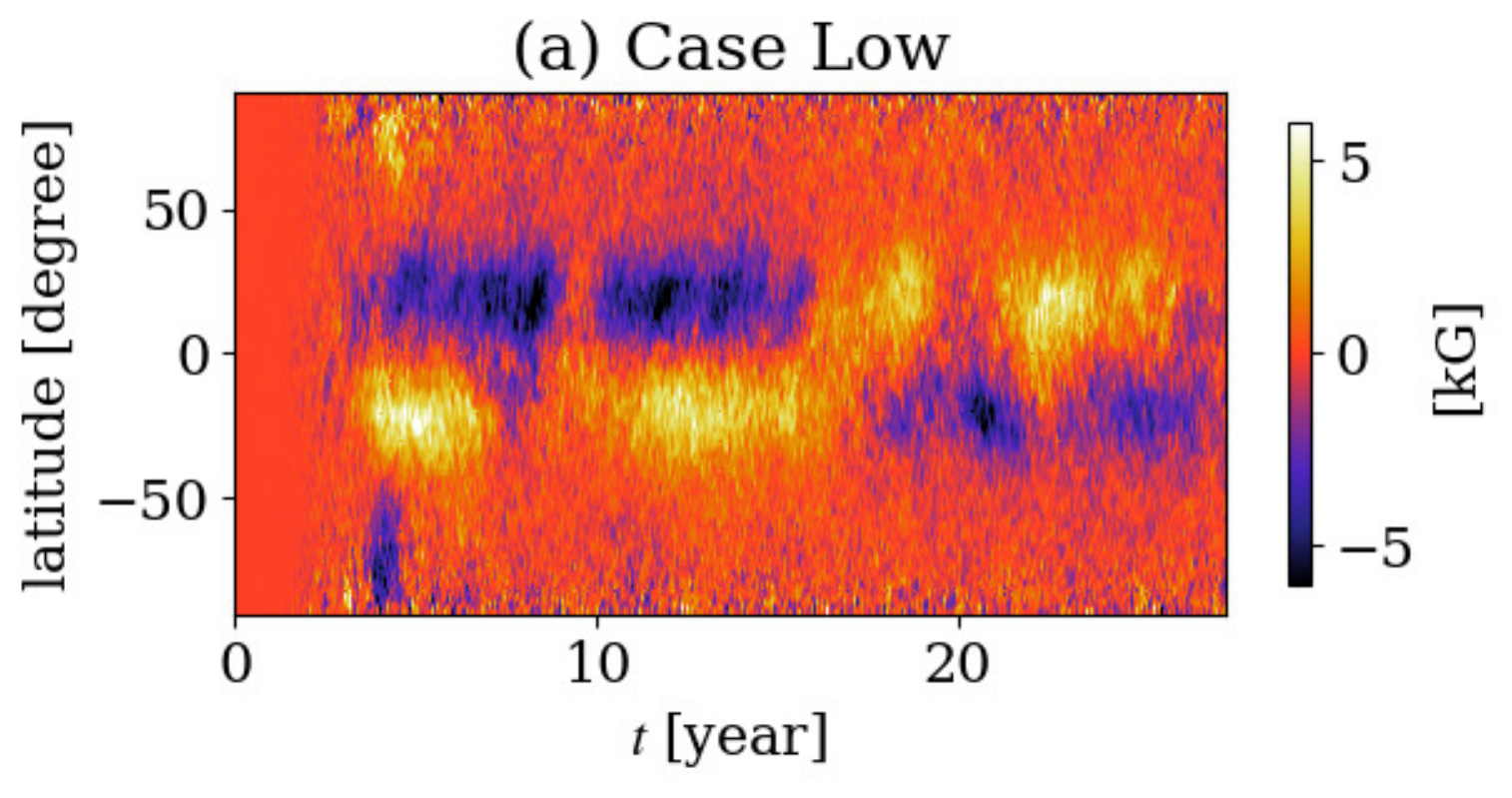}
  \plotone{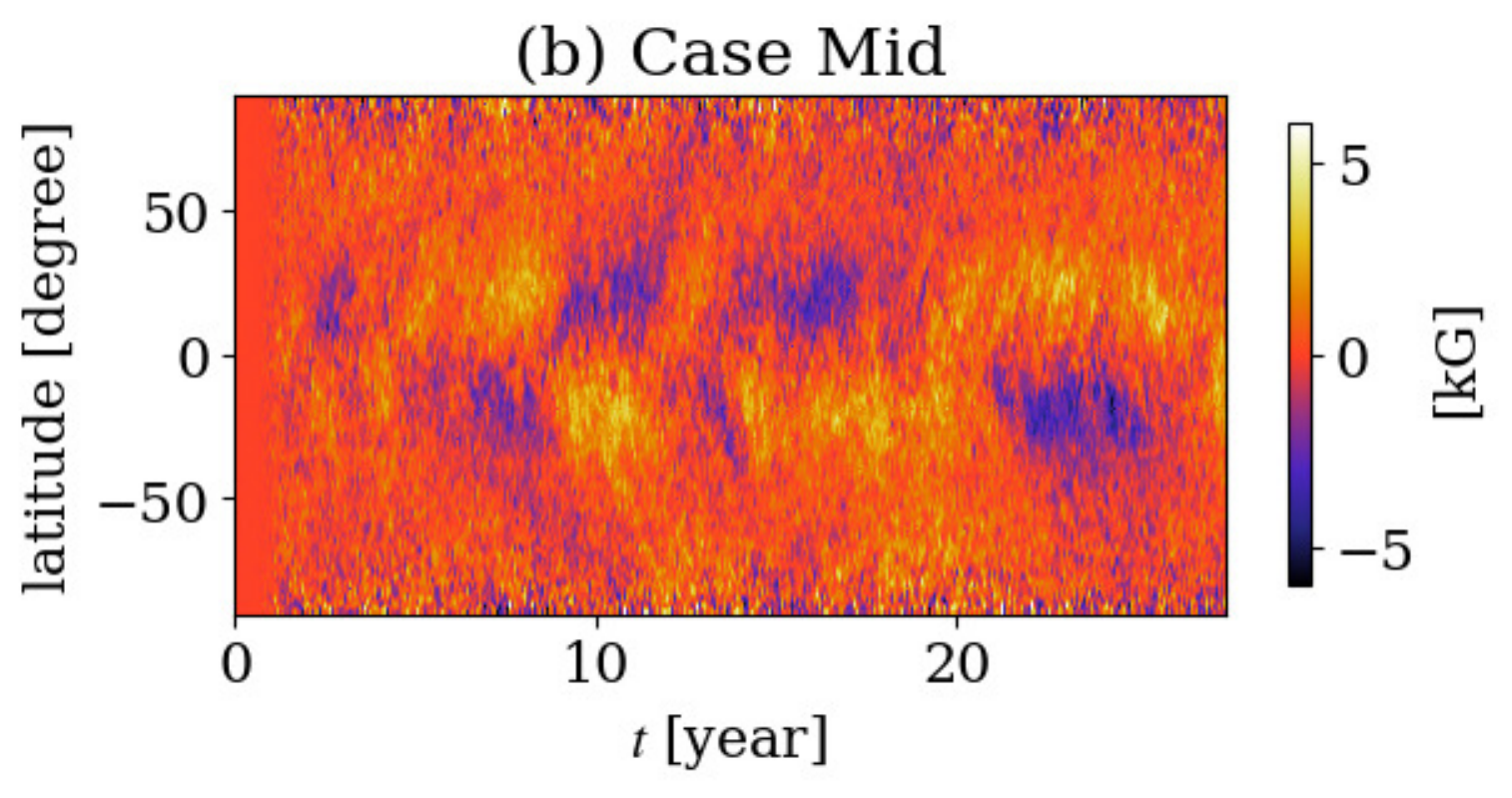}
  \plotone{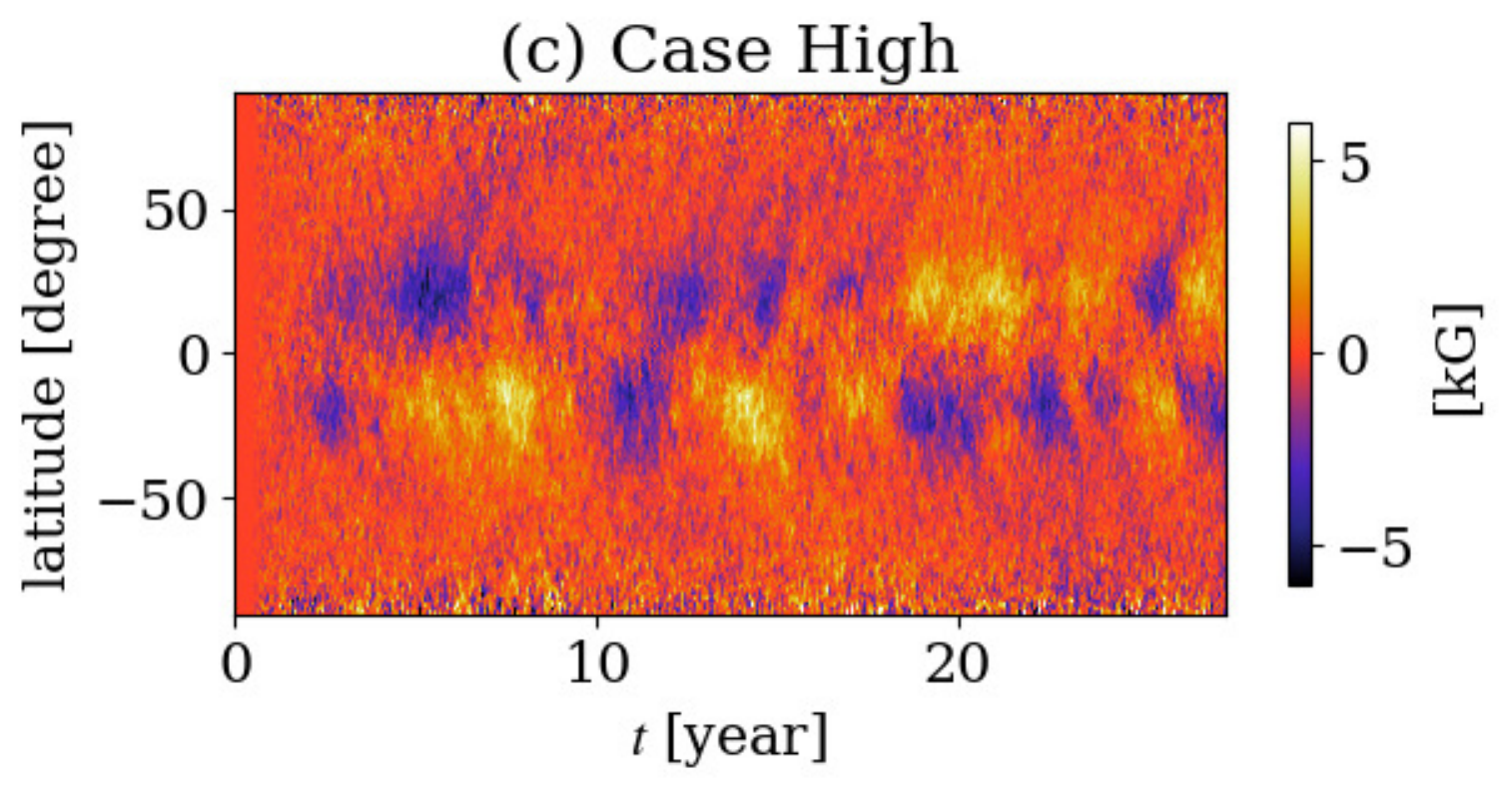}
  \caption{Temporal evolution of the radial distribution of $\braket{B_\varphi}$ at $r=0.72R_\odot$}
  \label{fig:butterfly}
\end{figure}

Figure \ref{fig:snapshot} presents the snapshots of radial velocity and longitudinal magnetic field in case High. 
Because the obtained large-scale magnetic field obviously has axisymmetric feature, we define mean-field component of given physical quantity $Q$ by:
\begin{eqnarray}
  &&\braket{Q}(t,r,\theta)=\frac{1}{2\pi}\int_0^{2\pi}d\varphi\ Q(t,r,\theta,\varphi)\label{eq:eq1}
\end{eqnarray} 
and fluctuating one by:
\begin{eqnarray}
  &&Q'(t,r,\theta,\varphi)=Q(t,r,\theta,\varphi)-\braket{Q}(t,r,\theta),\label{eq:eq2}
\end{eqnarray} 
for discussions.
Hereafter, $\braket{}$ and $'$ denote the operations described in Equations (\ref{eq:eq1}) and (\ref{eq:eq2}).

The large-scale magnetic field is concentrated around the base of the convection zone ($r<0.8R_\odot$) (see Figure \ref{fig:butterfly_2} in Appendix \ref{sec:appendix Large-Scale Field}). 
The temporal evolution of the mean toroidal magnetic field $\braket{B_\varphi}$ at the base of the convection zone is shown in Figure \ref{fig:butterfly}.
All the cases have spatially coherent structures at lower latitudes ($|\Theta|\la 30\arcdeg$, $\Theta=90\arcdeg-\theta$).
These cases also exhibit irregular polarity reversals of the large-scale magnetic field.
The reversal occurs every 5-10 yr in case Low and every 2-5 yr in the other cases. 
\begin{figure}
  \epsscale{1}
  \plotone{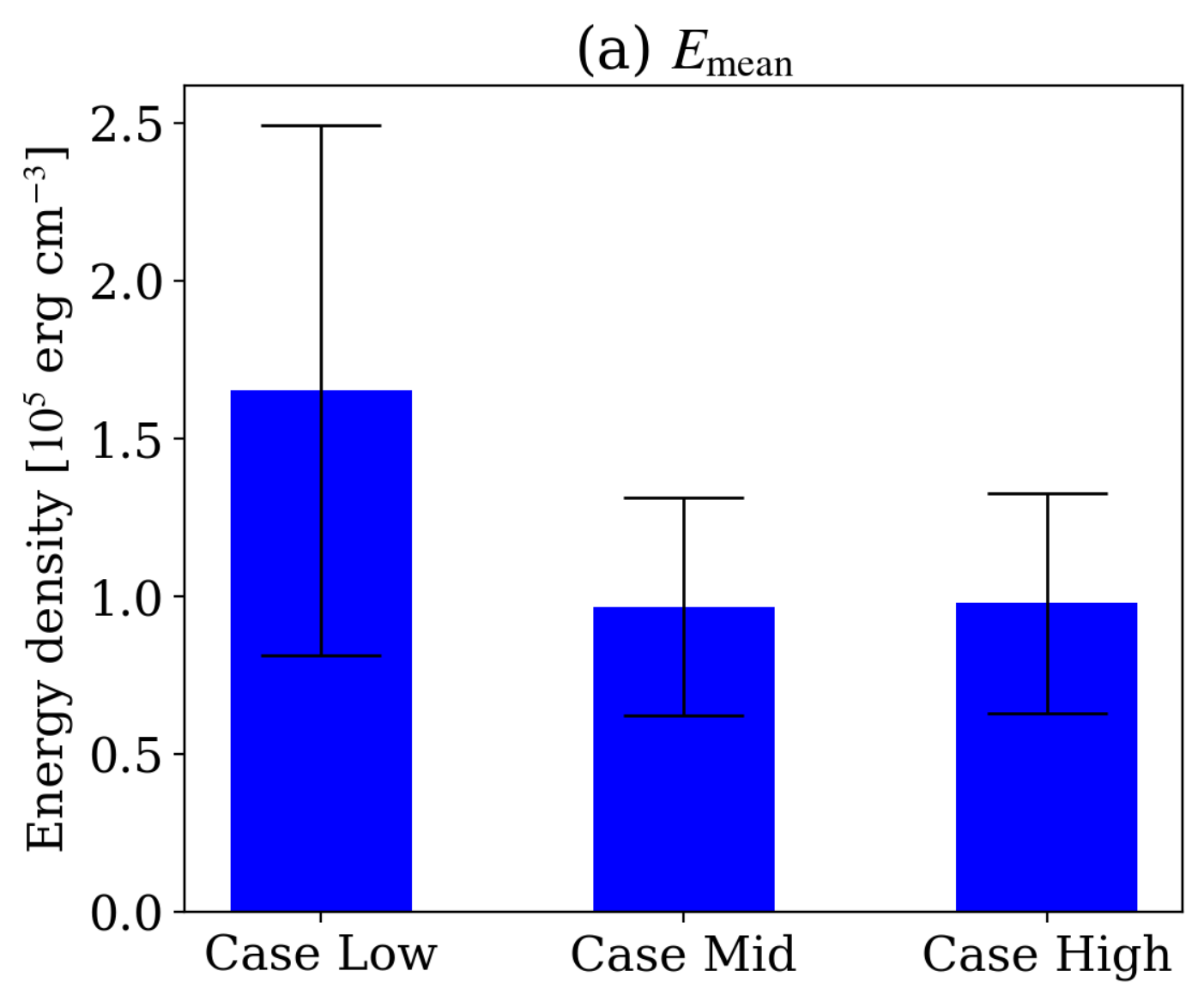}
  \plotone{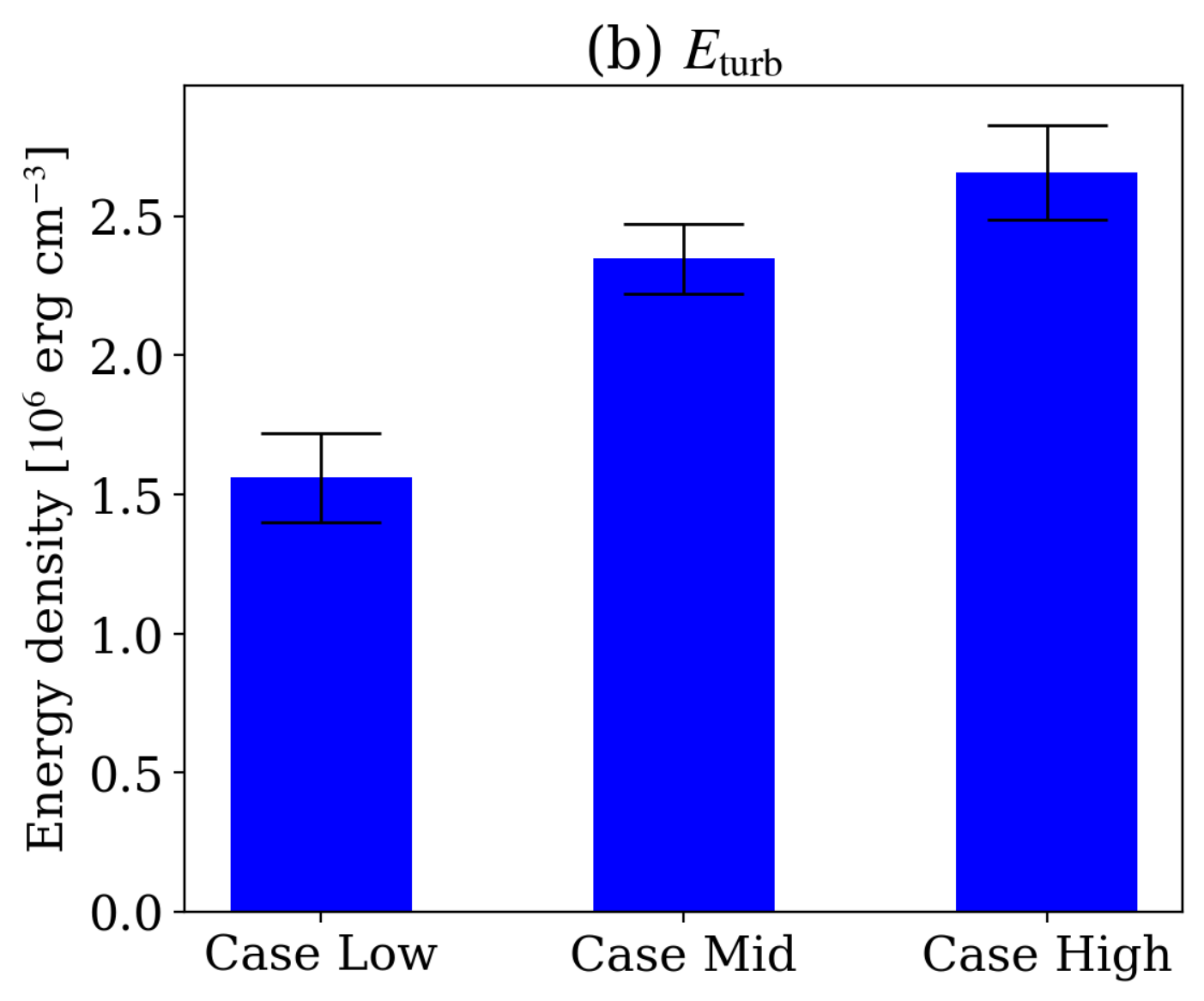}
  \caption{Energy densities of the large-scale magnetic field $E_{\mathrm{mean}}$ and the small-scale magnetic field $E_\mathrm{turb}$ at the base of the convection zone ($0.71R_\sun <r<0.73R_\sun$) are shown. $E_\mathrm{turb}$ is considered as the temporal average over a period of 5.5-27.4 yr, and $E_\mathrm{mean}$ is averaged over the period in which $E_\mathrm{mean}$ exceeds $5\times 10^4$ erg cm$^{-3}$ from 5.5-27.4 yr. Error bars indicate standard deviations.}
  \label{fig:bar}
\end{figure}

The strengths of the large-scale and small-scale magnetic fields depend on the cases.
The energy densities of the large-scale and small-scale magnetic fields are calculated as follows: 
\begin{eqnarray}
  \displaystyle E_\mathrm{mean}=\frac{1}{V}\int_V dV \frac{1}{8\pi}\braket{\bm{B}}^2,\label{eq:eq3}\\
  \displaystyle E_{\mathrm{turb}}=\frac{1}{V}\int_V dV \frac{1}{8\pi}\bm{B}'^2,\label{eq:eq4}
\end{eqnarray} 
where $V$ denotes the integrated volume. 
A volume localized at the base of the convection zone ($r<0.73R_\sun$) is used for $V$ in the following discussion.
Figure \ref{fig:bar} presents $E_\mathrm{mean}$ and $E_\mathrm{turb}$. 
The energy density of the large-scale magnetic field decreases from case Low to Mid and remains at the same level from case Mid to High.
Conversely, the energy density of the small-scale magnetic field monotonically increases toward case High with the increase in the magnetic Reynolds number.
$E_\mathrm{mean}$ is averaged over the period in which $E_\mathrm{mean}$ exceeds $5\times 10^4$ erg cm$^{-3}$ in 5.5 yr $< t <$27.4 yr.
The purpose of this operation is to avoid the misreading of the energy density of the large-scale magnetic field by including the period of polarity reversal. 
The large-scale magnetic field observed in case Mid is incoherent in time and does not vanish when the polarity is reversed, whereas that of cases Low and High is coherent in time and vanishes for a significant amount of time ($\backsimeq$1 yr) when the polarity is reversed (Figure \ref{fig:butterfly}).

\begin{figure}
  \gridline{\fig{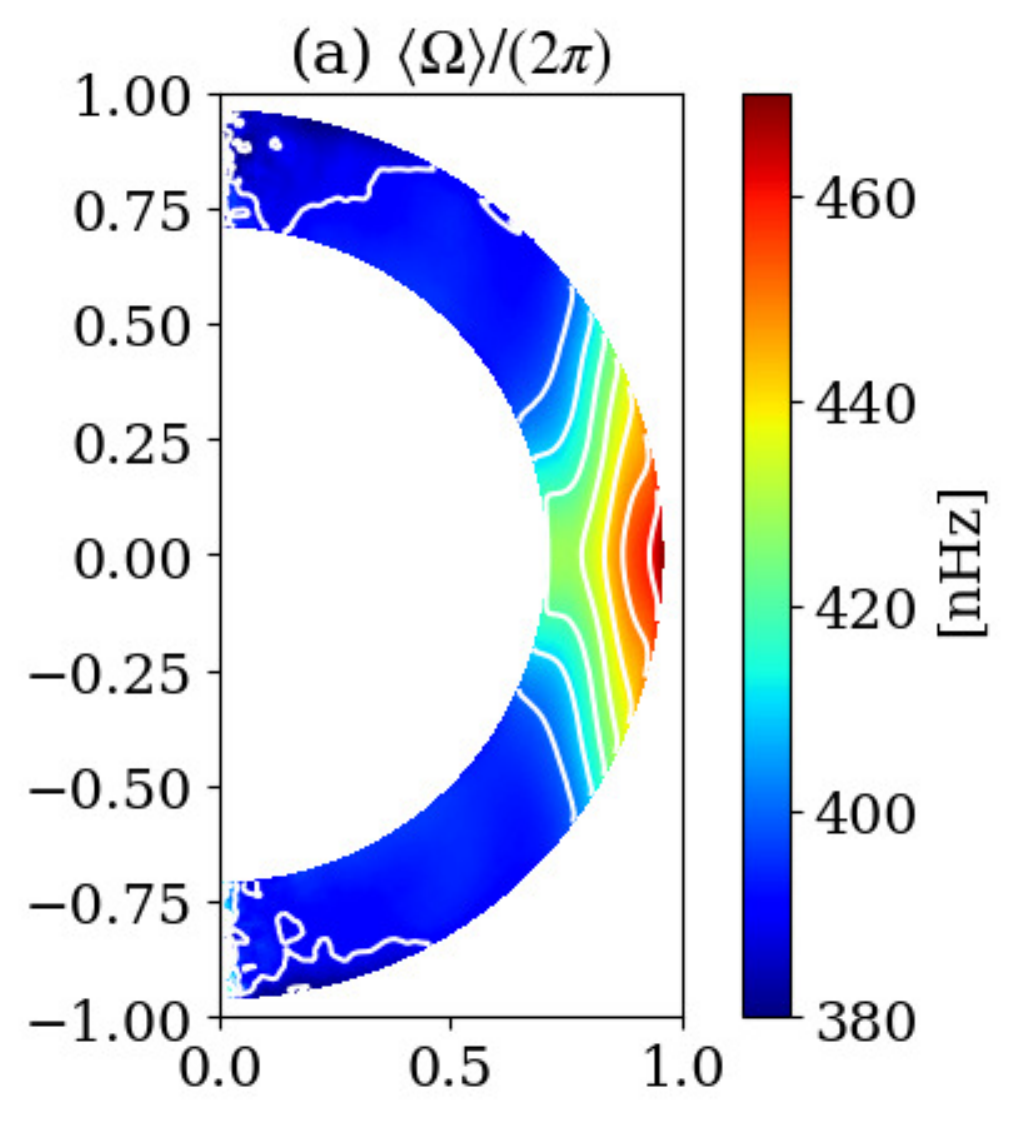}{0.24\textwidth}{}
            \fig{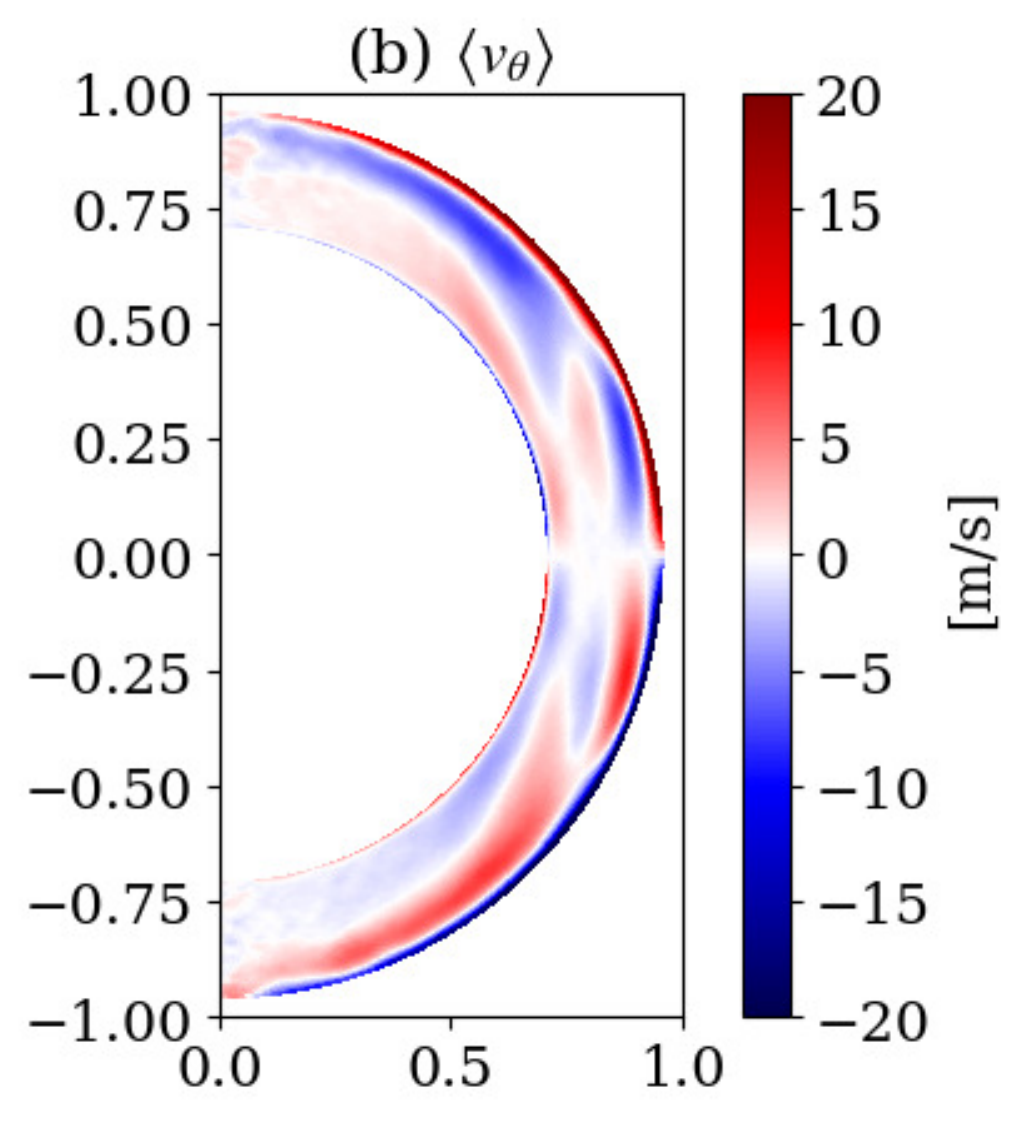}{0.24\textwidth}{}}
  \caption{(a) Differential rotation from case High is plotted on meridional plane. (b) Latitudinal mean flow $\braket{v_\theta}$ from case High is plotted on the meridional plane. }
  \label{fig:highflow}
\end{figure}

Figure \ref{fig:highflow} depicts the mean flow field.
It exhibits solar-like differential rotation such that the equator region rotates faster than the polar region.
The shear of the differential rotation decreases in case Mid and recovers in case High.
The mean meridional flow indicates the existence of two-cell circulation in each hemisphere and elongated structures parallel to the rotation axis at low latitudes.
The basic structures are almost identical in all the cases.
\begin{figure}
  \epsscale{1.2}
  \plotone{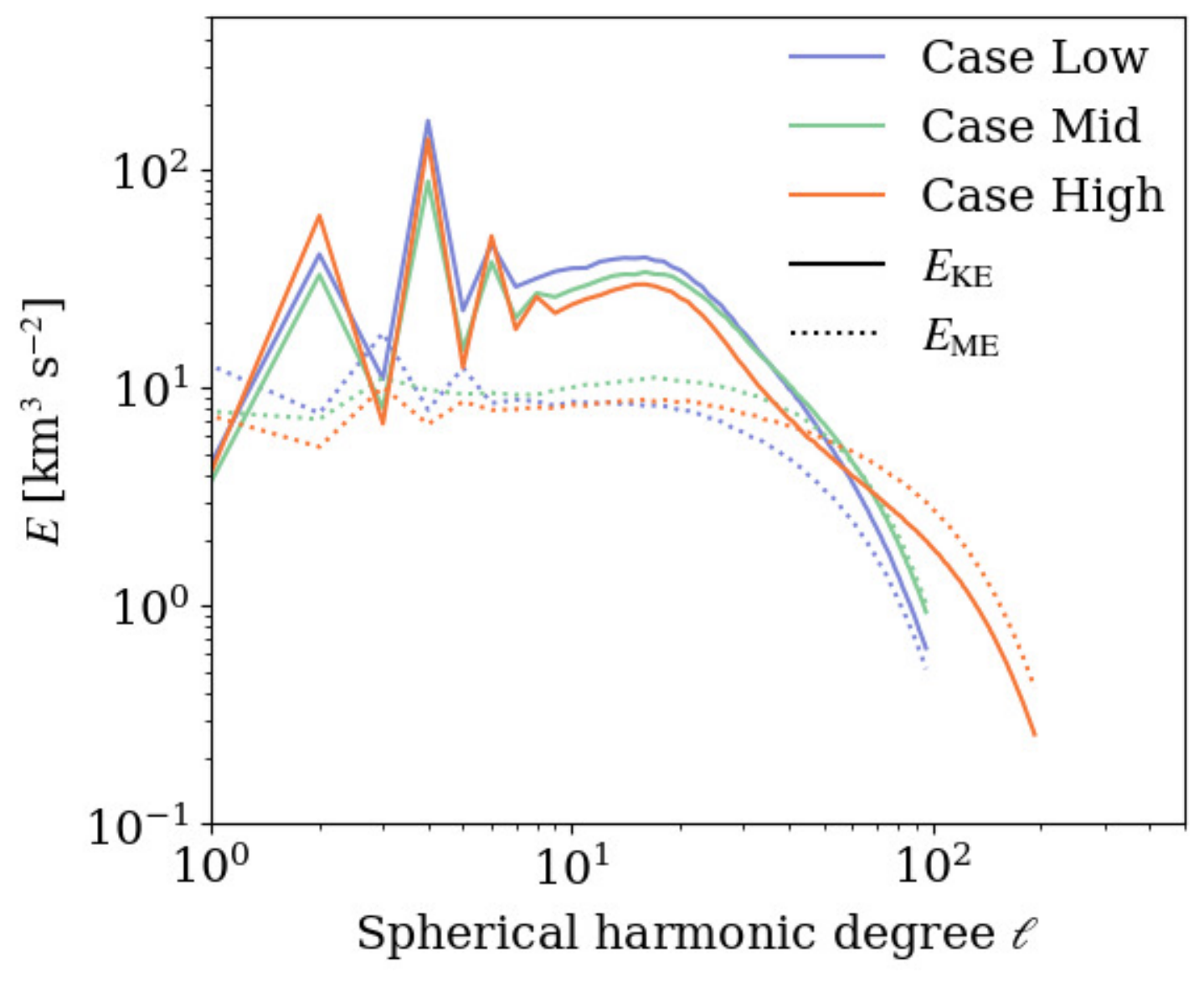}
  \caption{The kinetic (solid) and magnetic (dotted) energy spectra at $r=0.72R_\sun$. The blue, green, and orange lines represent the results from cases Low, Mid, and High, respectively. The averaged period is from 5.5 yr to 27.4 yr. }
  \label{fig:spectra}
\end{figure}   

The relative importance of the magnetic field and the flow field depends on the scale, and it can be measured using the energy spectra.
The kinetic energy spectra ($E_{\rm{KE}}(\ell)$) is compared with the magnetic energy spectra ($E_{\rm{ME}}(\ell)$). Our normalization satisfies the following relations:
\begin{eqnarray}
  &&\frac{1}{4\pi}\int_0^\pi d\theta\int_0^{2\pi} d\varphi\sin\theta\frac{\bm{v}^2}{2}=\sum_\ell E_{\rm{KE}}(\ell)/r,\label{eq:eq5}\\
  &&\frac{1}{4\pi}\int_0^\pi d\theta\int_0^{2\pi} d\varphi\sin\theta\frac{\bm{B}^2}{8\pi\rho_0}=\sum_\ell E_{\rm{ME}}(\ell)/r,\label{eq:eq6}
\end{eqnarray} 
where $\ell$ denotes the spherical harmonic degree, $\bm{v}$ and $\bm{B}$ represent both the mean-field and fluctuating components, and $\rho_0$ represents the density of the background stratification. 
Figure \ref{fig:spectra} depicts these quantities at $r=0.72R_\sun$ from cases Low, Mid, and High. 
For cases Low and Mid, the kinetic energy exceeds the magnetic energy in almost all the spatial scales, whereas the magnetic energy exceeds the kinetic energy at a smaller scale for case High.
This tendency in case High is also reported by \citet{2016Sci...351...1427}, in which they concluded that the small-scale dynamo efficiently suppresses the turbulent flow, and a large-scale magnetic field is realized as a result.

\section{MEAN-FIELD ANALYSIS} \label{sec:mean-field analysis}
The induction equation of the mean magnetic field is given by:
\begin{eqnarray}
  &&\frac{\partial \braket{\bm{B}}}{\partial t}=\nabla\times(\braket{\bm{v}}\times\braket{\bm{B}}+\bm{\mathcal{E}}-\eta\nabla\times\braket{\bm{B}}),\label{eq:eq7}
\end{eqnarray} 
where $\bm{\mathcal{E}}$ represents the so-called turbulent electromotive force (EMF), described as:
\begin{eqnarray}
  &&\bm{\mathcal{E}}=\braket{\bm{v}'\times\bm{B}'}.\label{eq:eq8}
\end{eqnarray} 
The EMF can be described by an expansion of the mean magnetic field component and its first derivatives as follows:
\begin{eqnarray}
  &&\bm{\mathcal{E}}=\bm{a}\cdot\braket{\bm{B}}+\bm{b}\cdot\nabla\braket{\bm{B}},\label{eq:eq22}
 \end{eqnarray} 
 where $\bm{a}$ denotes a rank-two tensor and $\bm{b}$ denotes a rank-three tensor. To obtain a more physically meaningful form, Equation (\ref{eq:eq22}) can be rewritten as:
\begin{eqnarray}
  \bm{\mathcal{E}}=&&\bm{\alpha}\cdot\braket{\bm{B}}+\bm{\gamma}\times\braket{\bm{B}}-\bm{\beta}\cdot(\nabla\times\braket{\bm{B}})\nonumber\\
  &&-\bm{\delta}\times(\nabla\times\braket{\bm{B}})-\bm{\kappa}\cdot(\nabla\braket{\bm{B}})^{(\mathrm{sym})},\label{eq:eq23}
\end{eqnarray}
where
\begin{eqnarray}
  &&\alpha_{ij}=\frac{1}{2}(a_{ij}+a_{ji}),\ \gamma_i=-\frac{1}{2}\epsilon_{ijk}a_{jk}\nonumber\\
  &&\beta_{ij}=-\frac{1}{4}(\epsilon_{ilm}b_{jlm}+\epsilon_{jlm}b_{ilm}),\nonumber\\
  &&\delta_i=\frac{1}{4}(b_{jij}-b_{jji}),\ \kappa_{ijk}=-\frac{1}{2}(b_{ijk}+b_{ikj})\label{eq:eq24}
\end{eqnarray}
\citep[see][]{2007_GApFD_101_81P}. 
$(\nabla\braket{\bm{B}})^{(\mathrm{sym})}$ represents the symmetric part of the gradient tensor, which is defined as:
\begin{eqnarray}
  &&(\nabla\braket{\bm{B}})^{(\mathrm{sym})}=\frac{1}{2}\{(\nabla\braket{\bm{B}})+(\nabla\braket{\bm{B}})^{\mathrm{T}}\}, \label{eq:eq50}
\end{eqnarray}
where $^{\mathrm{T}}$ is a transpose operation.
The signs of $\beta_{ij}$ and $\delta_{i}$ differ from those in \citet{2007_GApFD_101_81P}. 
$\bm{\alpha}$ and $\bm{\beta}$ are symmetric rank-two tensors, $\bm{\gamma}$ and $\bm{\delta}$ are vectors, and $\bm{\kappa}$ is a rank-three tensor. 
The $\alpha$-term is used to describe the $\alpha$-effect, which represents the contribution from helical flow.
$\bm{\gamma}$ is the virtual velocity known as turbulent pumping, which advects, shears, and compresses the mean-field as if it is the physical velocity.
The $\bm{\beta}$ term works as the turbulent diffusion.

\subsection{$\alpha$-tensor and $\beta$-tensor} \label{subsec:the alpha tensor}
The $\alpha$-tensor, $\gamma$-vector, and $\beta$-tensor are extracted from the 3D simulation results to understand mean-field induction. 
The method reported in \citet{2016_AdSpR_58_1522P} is employed in this procedure.
This method is an extension of \citet{2011ApJ...735..46P} to include the first derivatives of the mean field in the fitting procedure. 
It is based on a linear least-squares fit of the temporal variation of the EMF to that of the mean magnetic field component, along with the component of its first derivatives.
The relation between the EMF and the mean magnetic field, which is used for the linear least-squares fit, given by:
\begin{eqnarray}
  \mathcal{E}_i(t,r,\theta)=&&\tilde{a}_{ij}(r,\theta)\braket{B_j}(t,r,\theta)\nonumber\\
   &&+\tilde{b}_{ijr}(r,\theta)\frac{\partial \braket{B_j}}{\partial r}(t,r,\theta)\nonumber\\
   &&+\frac{\tilde{b}_{ij\theta}}{r}\frac{\partial \braket{B_j}}{\partial \theta},\label{eq:eq25}
\end{eqnarray} 
$\tilde{a}$ and $\tilde{b}$ are pseudo-tensors and are assumed to be time-independent.
The following procedures are performed at each grid point, $(r_b,\theta_c)$, and for each component, $(m=r,\theta$ or $\varphi)$, of the EMF.
We define 
\begin{eqnarray}
  y(t)=&&\mathcal{E}_m(t,r_b,\theta_c),\label{eq:eq10}\\
  \displaystyle X_k(t)=&&\left[\braket{B_k(t,r_b,\theta_c)},\right.\nonumber\\\
  &&\displaystyle \left.\frac{\partial \braket{B_k(t,r_b,\theta_c)}}{\partial r},\frac{1}{r}\frac{\partial \braket{B_k(t,r_b,\theta_c)}}{\partial \theta}\right],\label{eq:eq11}
\end{eqnarray}
and
\begin{eqnarray} 
  &&\Phi_k=\left[\tilde{a}_{mk}(r_b,\theta_c),\tilde{b}_{mkr}(r_b,\theta_c),\tilde{b}_{mk\theta}(r_b,\theta_c)\right],\label{eq:eq12}
\end{eqnarray} 
where $k=r, \theta, $ or $\varphi$.
Subsequently, the fitting formula for Equation (\ref{eq:eq25}) can be written as:
\begin{eqnarray}
  &&y(t)=\sum_{k}\Phi_kX_k(t).\label{eq:eq13}
\end{eqnarray}
A merit function $\chi^2$ is defined as:
\begin{eqnarray}
  &&\chi^2=\sum_{i=1}^{N_t}\left[y(t_i)-\sum_{k}\Phi_kX_k(t_i)\right]^2,\label{eq:eq14}
\end{eqnarray} 
where $N_t$ denotes number of time steps $t_i$ of the data. 
The best value for $\Phi_k$ is obtained by minimizing the merit function, $\chi^2$, using singular value decomposition (SVD).
The design matrix, $\bf{A}$, in the SVD is formed as:
\begin{eqnarray}
  &&A_{ij}=X_j(t_i).\label{eq:eq15}
\end{eqnarray} 
This matrix can be decomposed as:
\begin{eqnarray}
  &&\bf{A}=\bf{U}\cdot\bf{w}\cdot\bf{V}^{\rm{T}} ,\label{eq:eq16}
\end{eqnarray} 
where $\bf{U}$ is an $N_t\times9$ orthogonal matrix, $\bf{w}$ is a $9\times9$ diagonal matrix containing the singular values, and $\bf{V}$ is a $9\times9$ orthogonal matrix. 
The solution, $\bm{\Phi}=(\Phi_1,\cdots,\Phi_9)$, is given by:
\begin{eqnarray}
  &&\bm{\Phi}=\bf{V}\cdot\bf{w}^{-1}\cdot\bf{U}^{\rm{T}}\cdot\bm{y},\label{eq:eq17}
\end{eqnarray}
where $\Phi_1,\Phi_2,\Phi_3$ correspond to the $r,\theta,\varphi$ component of $\tilde{a}_{mk}$,
$\Phi_4,\Phi_5,\Phi_6$ correspond to the $r,\theta,\varphi$ component of $\tilde{b}_{mkr}$, 
and $\Phi_7,\Phi_8,\Phi_9$ correspond to the $r,\theta,\varphi$ component of $\tilde{b}_{mk\theta}$.
Each parameter, $\Phi_k$, includes the variance, $\sigma_k^2$, which is given by:
\begin{eqnarray}
  &&\sigma_k^2=\sigma^2\sum_{l=1}^9\left(\frac{V_{kl}}{w_{ll}}\right)^2,\label{eq:eq18}
\end{eqnarray}
where $\sigma^2$ denotes the variance of the merit function, and $V_{kl}$ and $w_{ll}$ represent the elements of the $\bf{V}$ and $\bf{w}$ matrices.
The relations between these pseudo-tensors ($\tilde{\bm{a}}$ and $\tilde{\bm{b}}$) and the true-tensors ($\bm{a}$ and $\bm{b}$) considering the curvilinear nature of the spherical coordinate system \citep[see][for further details]{2007_GApFD_101_81P}, are given as:
\begin{subeqnarray}
  &&a_{ir}=\tilde{a}_{ir}-\tilde{b}_{i\theta\theta}/r,\\
  &&a_{i\theta}=\tilde{a}_{i\theta}+\tilde{b}_{ir\theta}/r,\\
  &&a_{i\varphi}=\tilde{a}_{i\varphi},\\
  &&b_{ijr}=\tilde{b}_{ijr},\\
  &&b_{ij\theta}=\tilde{b}_{ij\theta},\\
  &&b_{ij\varphi}=0.\label{eq:eq26}
\end{subeqnarray} 
The $\alpha$-tensor, $\gamma$-vector, and $\beta$-tensor are automatically determined through Equations (\ref{eq:eq24}) and (\ref{eq:eq26}) after $\tilde{a}$ and $\tilde{b}$ are estimated by the fitting procedure. 
$\delta$-vector and $\kappa$-tensor are also obtained in this procedure.

\begin{figure}[!]
  \epsscale{1.2}
  \plotone{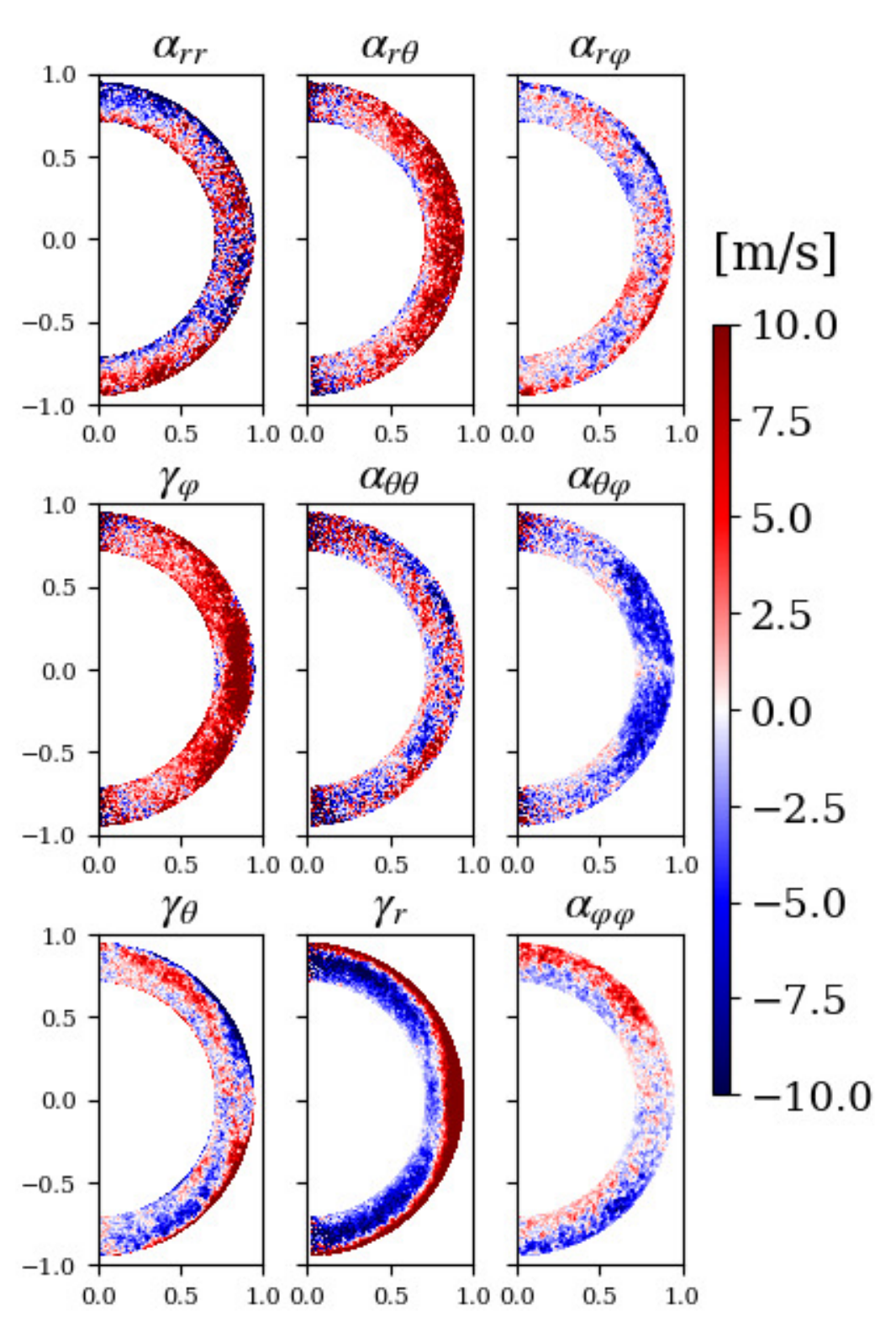}
  \caption{Components of the $a$-tensor extracted from case High plotted in meridional plane. The signs of $\gamma_\theta$, $\alpha_{r\theta}$, and $\alpha_{\theta\varphi}$ are different from those in \citet{2011ApJ...735..46P} because of the definition of the coordinates.}
  \label{fig:alpha}
\end{figure}

Firstly, the results of the $\alpha$-tensor and $\gamma$-vector obtained through this procedure are explained for each case using the entire simulation interval. 
Figure \ref{fig:alpha} presents the result from case High. 
The diagonal components, $\alpha_{rr}$, $\alpha_{\theta\theta}$, and $\alpha_{\varphi\varphi}$ are all antisymmetric about the equator and have different signs between the surface and the base.
Radial pumping, $\gamma_r$, is downward (negative) in most of the convection zone and upward (positive) in the subsurface. 
The latitudinal pumping, $\gamma_\theta$, is equatorward (positive/negative in the northern/southern hemisphere) in the lower convection zone and polarward (negative/positive in the northern/southern hemisphere) in the subsurface.
These properties are consistent with those of previous studies \citep{2011ApJ...735..46P, 2015_ApJ_809_149P,2016_AdSpR_58_1522P}.
It can be observed that the global structures of the $\alpha$-tensor are similar to those of the other cases. 
The resultant $\alpha$-tensor and $\gamma$-vector do not significantly vary from those estimated using the method reported in \citet{2011ApJ...735..46P}, which does not include the first derivatives of the mean magnetic field in the fitting procedure (see Appendix \ref{sec:appendix alpha tensor}).

\begin{figure}
  \epsscale{1.2}
  \plotone{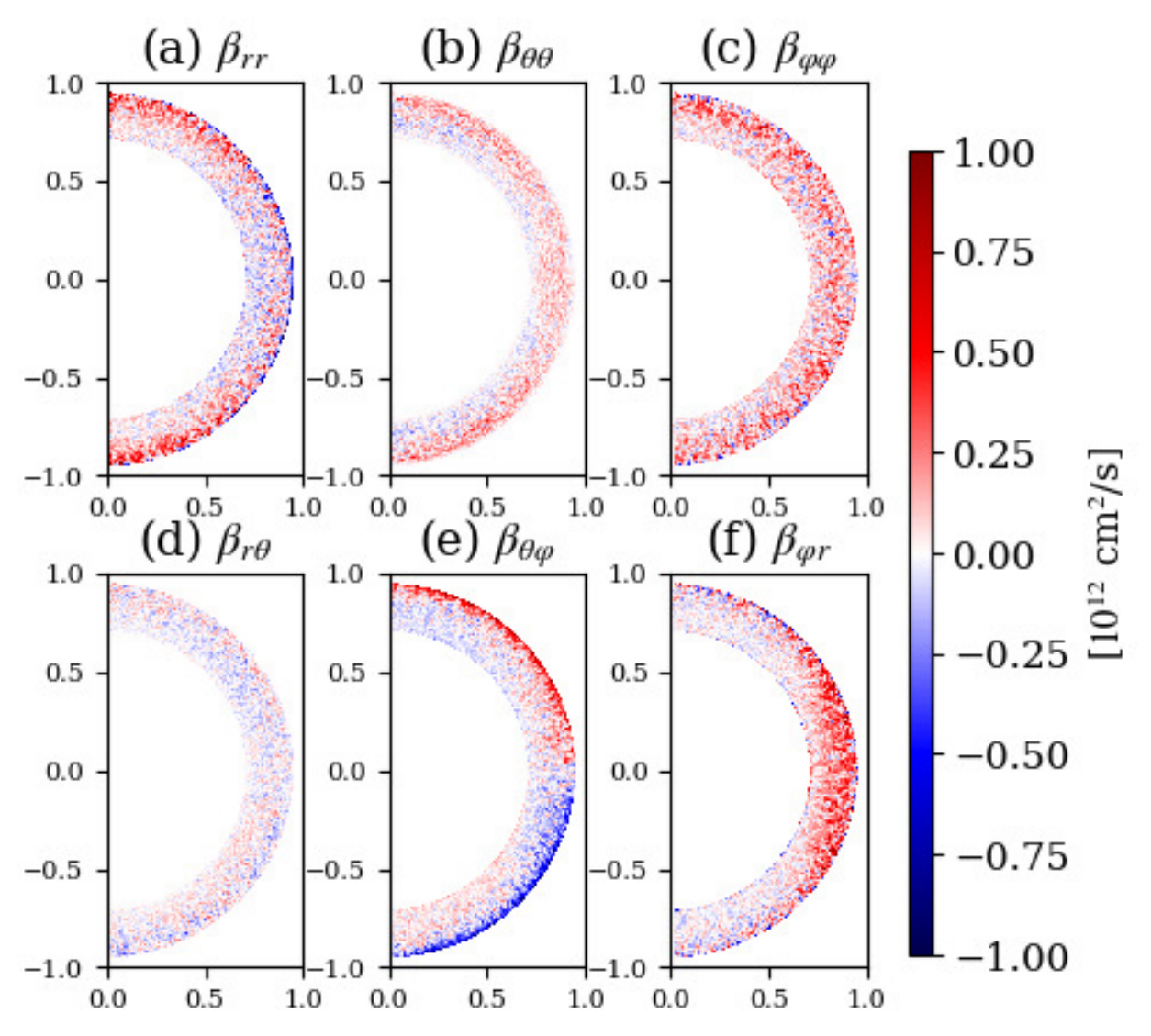}
  \caption{Components of the $\beta$-tensor extracted from case High plotted in meridional plane. }
  \label{fig:b_High}
\end{figure}

The importance measure, $\epsilon_{ij}$, introduced by \citet{2015_ApJ_809_149P}, is used to measure the relative importance of each component. 
It is defined as:
\begin{eqnarray}
 \epsilon_{ij}=&&\frac{3}{2E(r_2^3-r_1^3)}\nonumber\\
 &&\times\int_{r_1}^{r_2}dr\int_0^{\pi} d\theta\ r^2\sin\theta\sqrt{\frac{a_{ij}a_{ij}}{\{\bm{v}'\cdot\bm{v}'\}}},\label{eq:eq19}
 \end{eqnarray} 
where
\begin{eqnarray}
E=&&\frac{3}{2(r_2^3-r_1^3)}\nonumber\\
&&\times\sum_{i,j}\int_{r_1}^{r_2}dr\int_0^{\pi} d\theta\ r^2\sin\theta\sqrt{\frac{a_{ij}a_{ij}}{\{\bm{v}'\cdot\bm{v}'\}}},\label{eq:eq20}
 \end{eqnarray} 
 and
 \begin{eqnarray}
  \{\bm{v}'\cdot\bm{v}'\}=\sum_i\int dt \braket{v'_iv'_i}. \label{eq:eqn3}
   \end{eqnarray} 
$\{\bm{v}'\cdot\bm{v}'\}$ represents the sum of the diagonal elements of the Reynolds stress tensor averaged over the duration of the simulation and over all the longitudes. 
$\epsilon_{ij}\sim\Braket{\frac{a_{ij}}{v_\mathrm{rms}}}$, $E\sim\Braket{\frac{a}{v_{\mathrm{rms}}}}$, and $v_{\mathrm{rms}}\sim\braket{v_i'v_i'}$ where $a$ is a norm of $a$-tensor, $v_{\mathrm{rms}}$ is the root mean squared velocity.
$r_1$ and $r_2$ denote the radii of the lower and upper boundaries of the simulation domain, respectively. 
Importance measure, $\epsilon_{ij}$, from case High is calculated as:
 \begin{eqnarray}
 &&\left(\begin{array}{ccc} \epsilon_{\alpha_{rr}}& \epsilon_{\alpha_{r\theta}}  &\epsilon_{\alpha_{r\varphi}}\\  \epsilon_{\gamma_{\varphi}}& \epsilon_{\alpha_{\theta\theta}}&\epsilon_{\alpha_{\theta\varphi}}\\\epsilon_{\gamma_{\theta}}&\epsilon_{\gamma_{r}}&\epsilon_{\alpha_{\varphi\varphi}}\end{array}\right)
 =\left(
\begin{array}{ccc}
0.15&\	0.14&\	0.06\\
0.18&\	0.09&\	0.08\\
0.08&\	0.17&\	0.04
\end{array}     \right).\label{eq:eq21}
\end{eqnarray} 
This indicates that the upper two-by-two matrix formed by $\alpha_{rr}$, $\alpha_{r\theta}$, $\alpha_{\theta\theta}$, and $\gamma_{\varphi}$ are dominant over the others, which is consistent with \citet{2015_ApJ_809_149P}. 
However, our results exhibit a larger contribution than those of $\gamma_r$, which exceeds $\alpha_{rr}$, $\alpha_{\theta\theta}$, and $\alpha_{r\theta}$. 
In their study, the importance measures take the value, $0.054$, in $\epsilon_{\gamma_r}$, $0.355$ in $\epsilon_{\alpha_{rr}}$, $0.103$ in $\epsilon_{\alpha_{\theta\theta}}$, and $0.124$ in $\epsilon_{\alpha_{r\theta}}$.
The relative contribution of each component of the $a$-tensor does not vary significantly in all the cases.

\citet{warnecke_2018} extract the $\alpha$-tensor and $\gamma$-vector based on test-field methods. 
These quantities in our result (Figure \ref{fig:alpha}) have smaller structure than those in \citet{warnecke_2018}.
The amplitudes of the structures, whose scales are smaller than $0.01 R_\sun$ are within the range of 1$\sigma$ and only the structures larger than $0.01 R_\sun$ are reliable. 
$\gamma_r$ extracted by \citet{warnecke_2018} changes the sign between inside and outside the tangential cylinder, whereas our $\gamma_r$ has the same sign between these two regions. 
The cause of this difference might come from the difference in rotational constraint on the simulation, because the rotation rate of the simulation analyzed in \citet{warnecke_2018} is five times larger than that of ours.

\begin{figure}
  \epsscale{1.2}
  \plotone{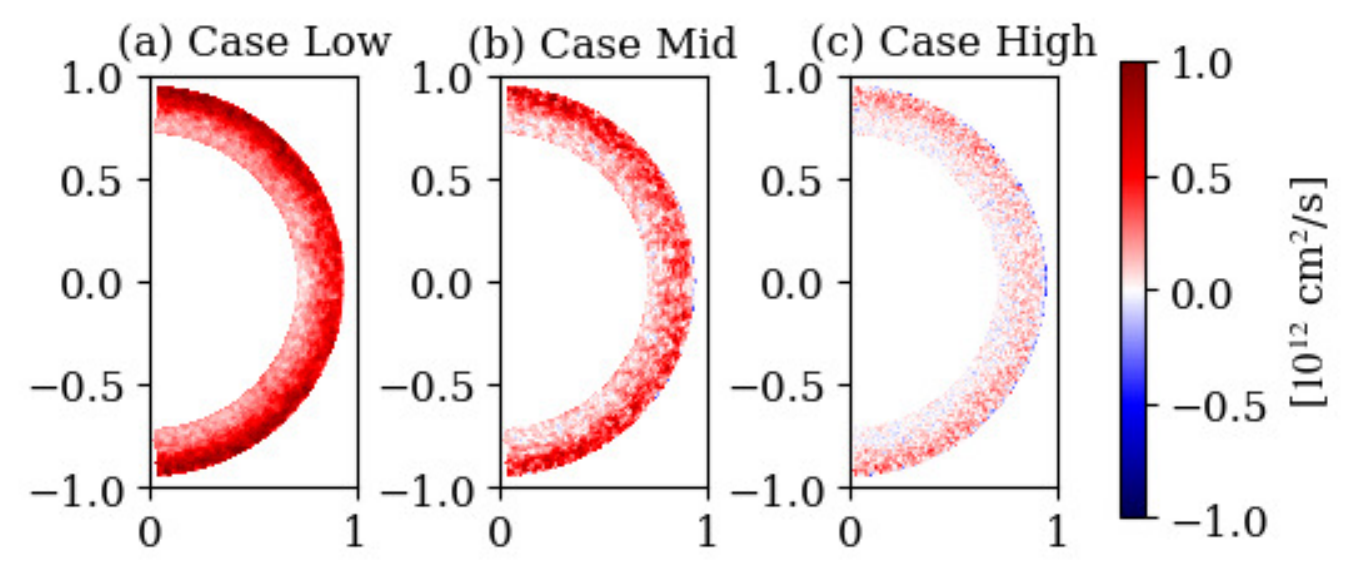}
  \caption{The isotropic part of the $\beta$-tensor ($\beta$) extracted from cases Low, Mid, and High plotted in meridional plane.}
\label{fig:beta_3}  
\end{figure}

\begin{figure}
  \epsscale{1}
  \plotone{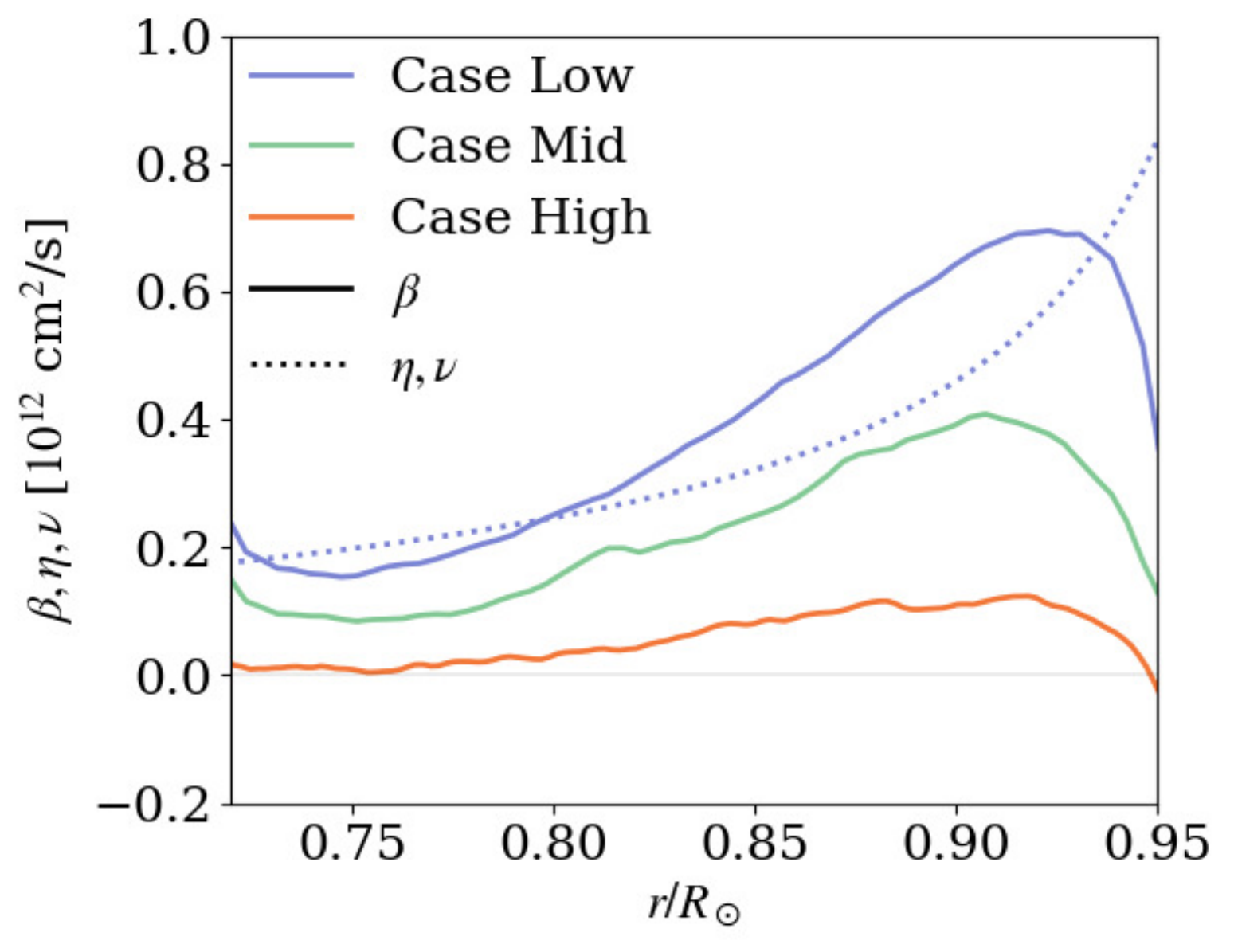}
  \caption{Radial plot of the turbulent magnetic diffusivity, $\beta$, explicit magnetic diffusivity, $\eta$, and explicit viscosity, $\nu$, averaged over the latitude from cases Low, Mid, and High.}
  \label{fig:beta_comp}
\end{figure}

Secondly, the results of the $\beta$-tensor obtained through this procedure are explained for each case using the entire simulation interval. 
Figure \ref{fig:b_High} shows the $\beta$-tensor extracted from case High. 
Negative values observed in the diagonal elements are within the range of 2$\sigma$.
Note that estimation of the $\alpha$-tensor is not significantly affected by the inclusion of the $\beta$-tensor (see Appendix \ref{sec:appendix alpha tensor}). 
To obtain the physical meaning from this $\beta$-tensor, the sum of the diagonal elements is calculated as: 
\begin{eqnarray}
  &&\beta\equiv\frac{1}{3}(\beta_{rr}+\beta_{\theta\theta}+\beta_{\varphi\varphi}),\label{eq:eq27}
\end{eqnarray}
where $\beta$ corresponds to the turbulent magnetic diffusivity introduced in the widely known mean-field dynamo model. 
Figure \ref{fig:beta_3} presents the spatial structure of $\beta$ from case Low to High. The result indicates that a positive value is observed through most of the convection zone and increases toward the surface. 
Figure \ref{fig:beta_comp} presents the radial plots of $\beta$ averaged over the latitudes. $\beta$ reaches its maximum value of 7$\times$10$^{11}$ cm$^2$ s$^{-1}$ for the Low, 4$\times10^{11}$ cm$^2$ s$^{-1}$ of case Mid, and 1$\times$10$^{11}$ cm$^2$ s$^{-1}$ of case High at depth $r/R_\sun\sim0.91$. 
It is clear that the turbulent magnetic diffusivity decreases with an increase in the magnetic Reynolds number.

Note that we omit the results of $\delta$-vector and $\kappa$-tensor because the values of some components have less than 1$\sigma$ noise level. 
Unfortunately, we can not compare these results with those in the previous work by \citet{2016_AdSpR_58_1522P}, because there were no description of the results nor related discussion while these vector and tensor were shown in the fitting formula (see their equation 5).

\subsection{Induction of the Mean Magnetic Field}\label{subsec:induction of the mean magnetic field}

\begin{figure}
  \epsscale{1.2}
  \plotone{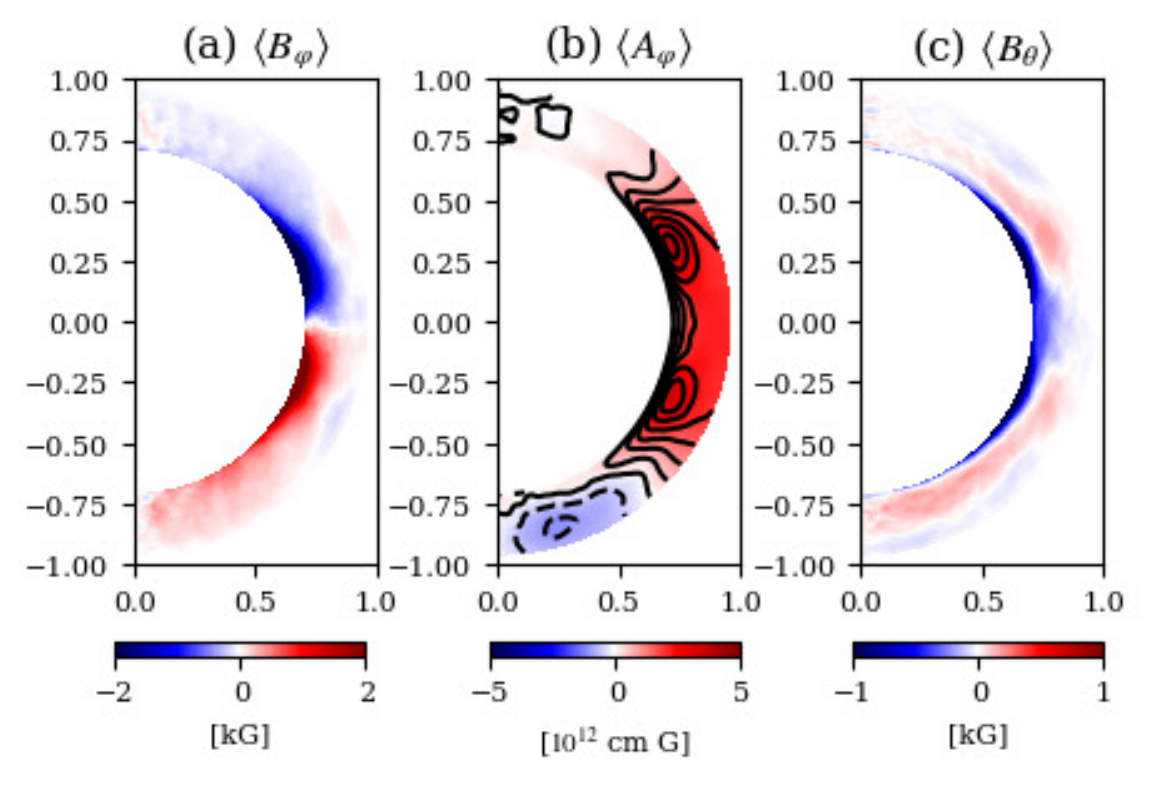}
  \caption{(a) $\braket{B_\varphi}$, (b) $\braket{A_\varphi}$, and (c) $\braket{B_\theta}$ from case Low plotted in meridional plane. The contours in (b) are at every 5 $\times$ 10$^{11}$ cm G.
  }
  \label{fig:average_B}
\end{figure}

A coherent large-scale magnetic field is observed in all the cases (see Figure \ref{fig:butterfly} and Subsection \ref{subsec:large-scale field}). 
The elemental processes required to maintain the mean field are analyzed to determine the construction mechanism of the large-scale magnetic field.
The induction equation can be decomposed as follows: \citep{2010_ApJ_711_424P,2012_ApJ_762_73P,2015_ApJ_809_149P}: 
\begin{eqnarray}
  \frac{\partial \braket{\bm{B}}}{\partial t}=&&\underbrace{(\braket{\bm{B}}\cdot\nabla)\braket{\bm{v}}}_{\mathrm{MS}}+\underbrace{\braket{(\bm{B}'\cdot\nabla)\bm{v}'}}_{\mathrm{FS}}\nonumber\\
  &&\underbrace{-(\braket{\bm{v}}\cdot\nabla)\braket{\bm{B}}}_{\mathrm{MA}}\underbrace{-\braket{(\bm{v}'\cdot\nabla)\bm{B}'}}_{\mathrm{FA}}\nonumber\\
  &&\underbrace{-\braket{\bm{B}}(\nabla\cdot\braket{\bm{v}})}_{\mathrm{MC}}\underbrace{-\braket{\bm{B}'(\nabla\cdot\bm{v}')}}_{\mathrm{FC}}\nonumber\\
  &&\underbrace{-\nabla\times(\eta\nabla\times\braket{\bm{B}})}_{\mathrm{RD}}.\label{eq:eq28}
\end{eqnarray} 
Equation (\ref{eq:eq28}) is used to search for important terms to induce the mean field. 
The terms on the right-hand side of Equation (\ref{eq:eq28}) represent the production of the mean magnetic field by the mean shear (MS), fluctuating shear (FS), mean advection (MA), fluctuating advection (FA), mean compression (MC), fluctuating compression (FC), and resistive diffusion (RD). 
In our results, the spatial distributions of these terms do not vary significantly from the cases, but some fluctuating structures that have a much smaller scale than the large-scale magnetic field are found in cases Mid and High. 
The result from case Low is presented in the following section to discuss the large-scale distribution.

\begin{figure}
  \epsscale{0.9}
  \plotone{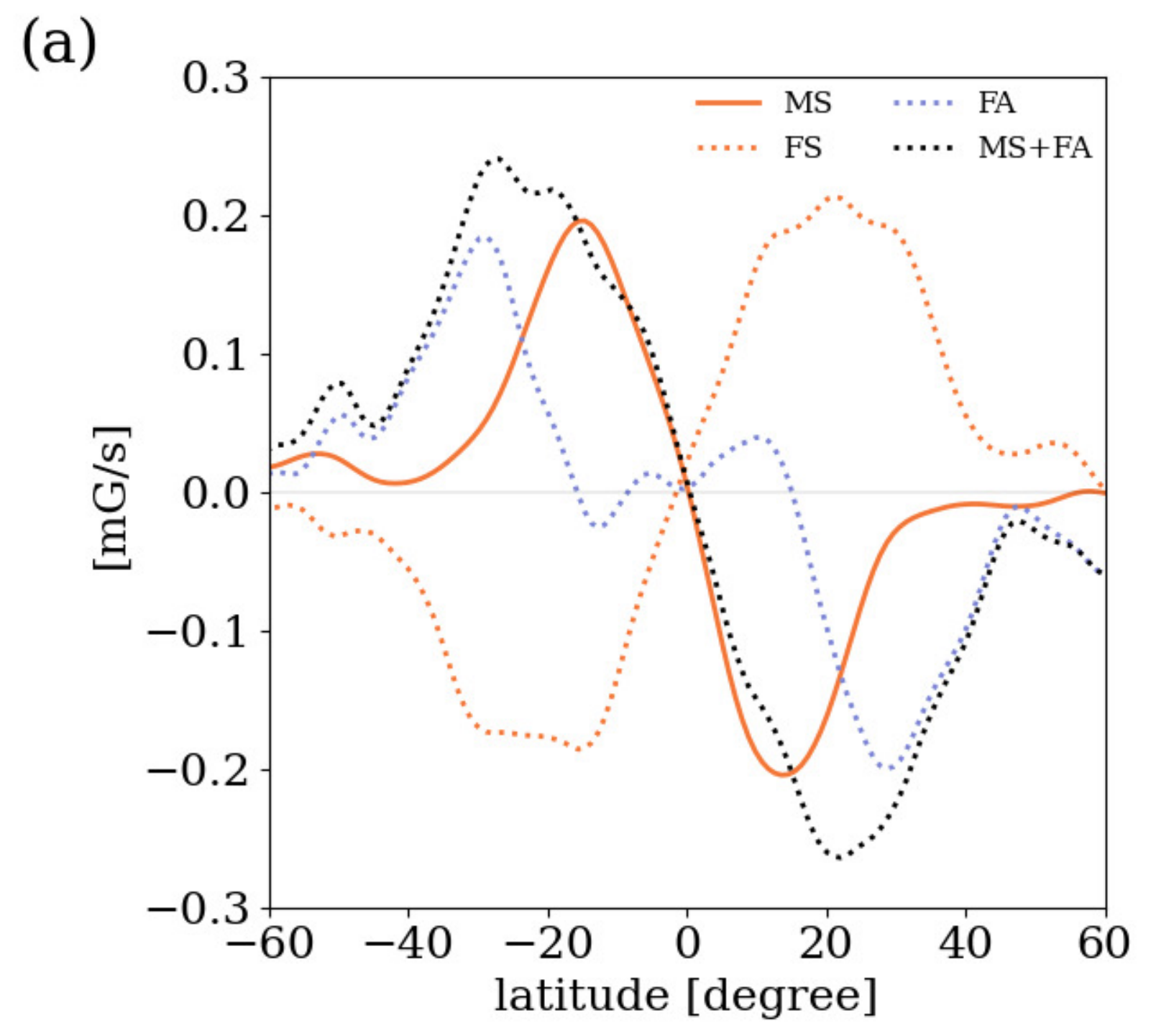}
  \plotone{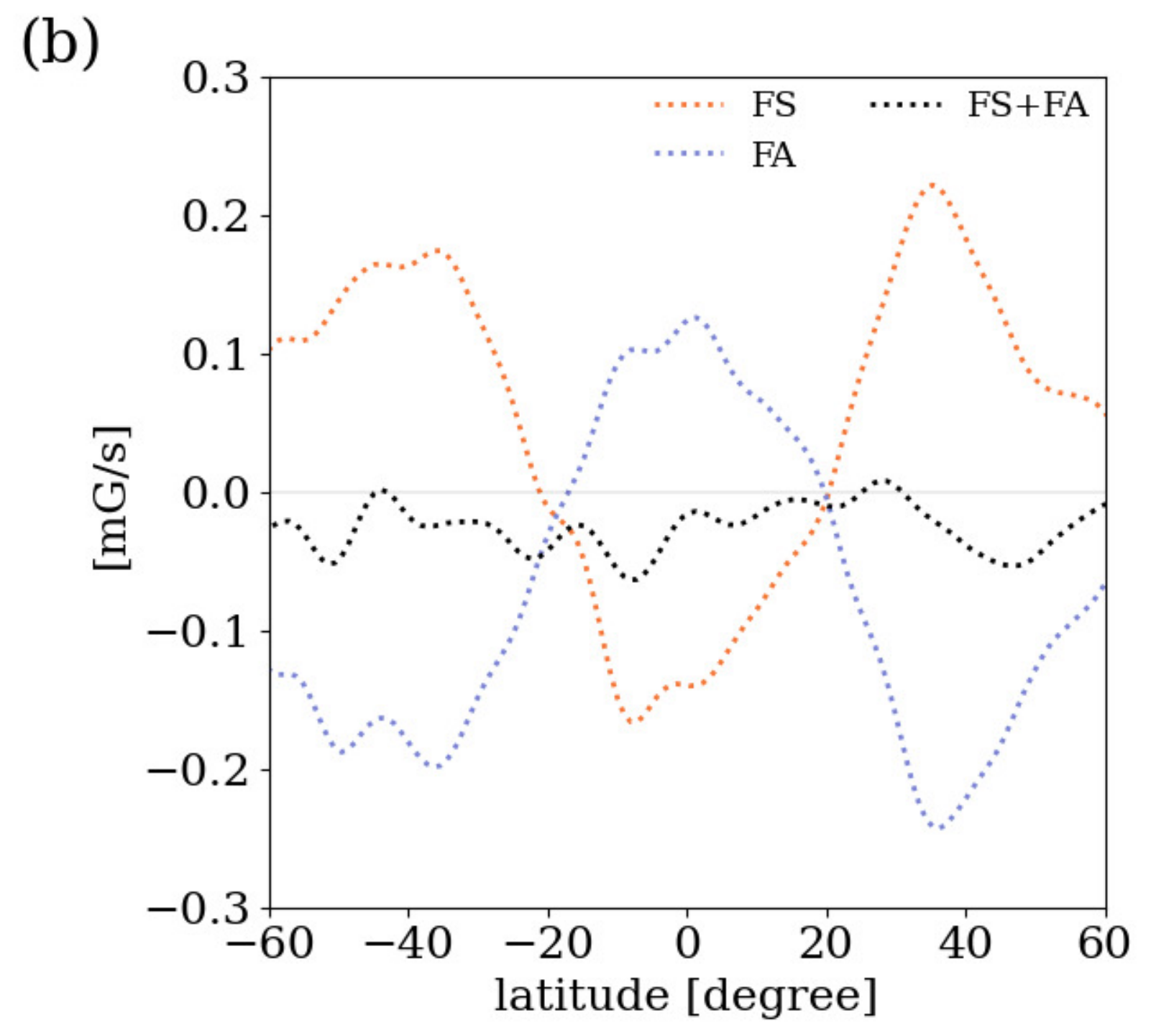}
  \caption{(a) Latitudinal plot of time-averaged the dominant terms in Equation (\ref{eq:eq28}) which contribute to the induction of $\braket{B_\varphi}$ at the base of convection zone ($r/R_{\sun}\leqq0.8$). (b) The same for $\braket{B_\theta}$. The results for case Low are presented. These are averaged over $0.71R_\sun\leqq r \leqq 0.80R_\sun$ and filtered by a Gaussian filter with a half-width of $0.01R_\odot$ in the latitudinal direction to read large-scale characteristics.}
  \label{fig:MS_FA}
\end{figure}

A time period stretching over approximately 5 yr, is selected to analyze the coherent large-scale magnetic field, when the mean magnetic field has constant polarity. 
The average was taken over this period.
Figure \ref{fig:average_B} presents the mean magnetic field of case Low.
The mean toroidal magnetic field, $\braket{B_\varphi}$, is concentrated at the base of the convection zone ($r/R_{\odot}\la0.8$) and low latitudes ($|\Theta|\la 30\arcdeg$). 
The toroidal magnetic vector potential, $\braket{A_\varphi}$, depicted in Figure \ref{fig:average_B} (b) captures the mean poloidal magnetic field as $\braket{B_r}\bm{e}_r+\braket{B_\theta}\bm{e}_\theta=\braket{\bm{B}_\mathrm{P}}=\nabla\times(\braket{A_\varphi}\bm{e}_\varphi)$. 
$\braket{B_\theta}$ is concentrated at the base of the convection zone ($r/R_{\odot}\la0.8$) and low latitudes ($|\Theta|\la 30\arcdeg$). 
$\braket{B_\theta}$ pointing in the opposite direction to the bottom one is distributed at the middle depth of the convection zone ($0.8\la r/R_{\odot}\la0.9$). 
These spatial structures of the mean magnetic field are common in the other cases, despite the difference in the amplitudes.

Figure \ref{fig:MS_FA} (a) presents the latitudinal distribution of the dominant terms in Equation (\ref{eq:eq28}), which contribute to the induction of $\braket{B_\varphi}$. 
These results are averaged over $0.71R_\sun\leqq r \leqq 0.80R_\sun$.
The contributions of all the terms are listed in Appendix \ref{sec:appendix whole terms of the induction equation}. 
The $\Omega$-effect (MS) and fluctuating advection (FA) produce $\braket{B_\varphi}$ at low latitudes ($|\Theta|\la 40\arcdeg$); the fluctuating shear (FS) works against production. 
The tendency of MS and FA to produce $\braket{B_\varphi}$ against FS is also observed in the other cases.
This is consistent with \citet{2012_ApJ_762_73P} for their case D3b.

Figure \ref{fig:MS_FA} (b) presents the latitudinal distribution of the dominant terms in Equation (\ref{eq:eq28}), which contributes to the induction of $\braket{B_\theta}$ at the base of the convection zone. 
These results are averaged over $0.71R_\sun\leqq r \leqq 0.80R_\sun$.
The contributions of all the terms are listed in Appendix \ref{sec:appendix whole terms of the induction equation}. 
Figure \ref{fig:MS_FA} (b) demonstrates that the fluctuating shear produces negative $\braket{B_\theta}$ at low latitudes ($|\Theta|\la 20\arcdeg$) against the fluctuating advection.
At high latitudes ($|\Theta|\ga 20\arcdeg$), positive $\braket{B_\theta}$ is observed around the base of the convection zone ($r/R_{\sun}\leqq0.8$), and the result presented in Figure \ref{fig:MS_FA} (b) is the radial average over $r/R_\odot \leq 0.8$, due to which the signs of these terms are reversed against the signs at low latitudes. 

\begin{figure}
  \epsscale{1.2}
  \plotone{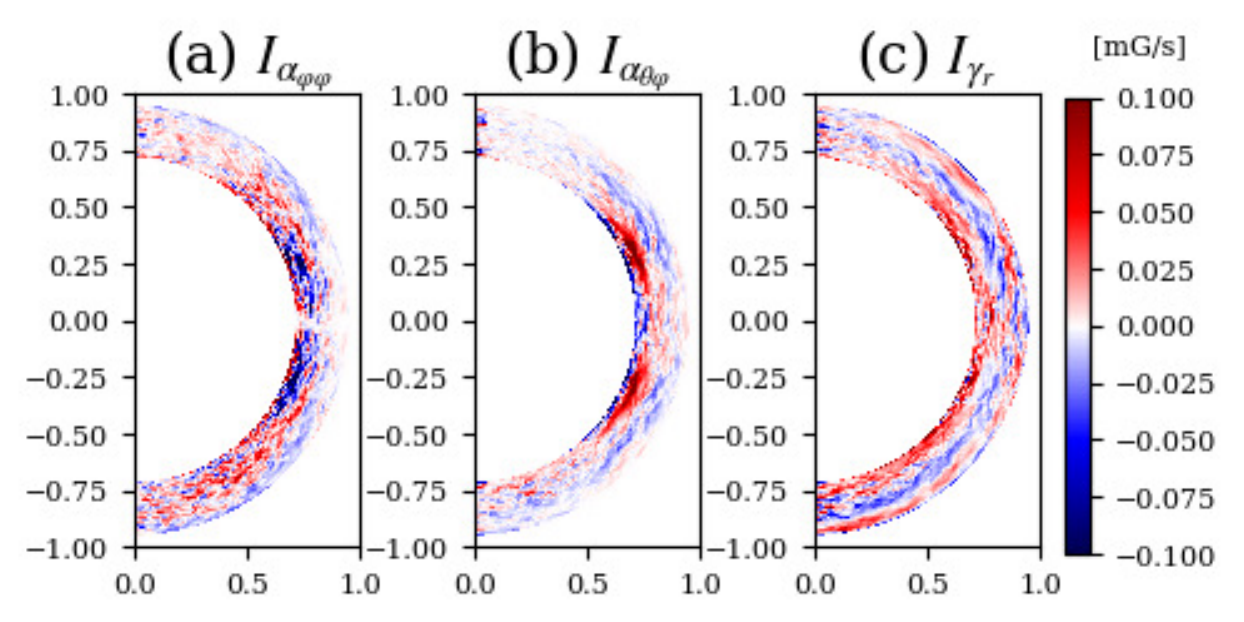}
  \caption{Contribution from (a) $\alpha_{\varphi\varphi}$, (b) $\alpha_{\theta\varphi}$, and (c) $\gamma_r$ for case Low are plotted in meridional plane. 
  }
  \label{fig:induce}
  \end{figure} 

The contribution of the fluctuating field, that is, EMF, is important in inducing $\braket{B_\theta}$ at the base of the convection zone and low latitudes, as explained earlier.
The EMF is decomposed by each component of the $\alpha$-tensor and $\gamma$-vector and their contribution is evaluated as: 
\begin{eqnarray}
  \{\nabla\times (\bm{a}\cdot\braket{\bm{B}})\}_\theta=&&\underbrace{-\frac{1}{r}\frac{\partial }{\partial r}(r\alpha_{r\varphi}\braket{B_r})}_{I_{\alpha_{r\varphi}}}+\underbrace{\frac{1}{r}\frac{\partial }{\partial r}(r\gamma_\theta \braket{B_r})}_{I_{\gamma_\theta}}\nonumber\\
  &&\underbrace{-\frac{1}{r}\frac{\partial }{\partial r}(r\alpha_{\theta\varphi}\braket{B_\theta})}_{I_{\alpha_{\theta\varphi}}}\underbrace{-\frac{1}{r}\frac{\partial }{\partial r}(r\gamma_r \braket{B_\theta})}_{I_{\gamma_r}}\nonumber\\
  &&\underbrace{-\frac{1}{r}\frac{\partial }{\partial r}(r\alpha_{\varphi\varphi}\braket{B_\varphi})}_{I_{\alpha_{\varphi\varphi}}}.
\end{eqnarray} 
Figure \ref{fig:induce} presents the dominant terms, that is, $I_{\alpha_{\varphi\varphi}}$, $I_{\alpha_{\theta\varphi}}$, and $I_{\gamma_r}$, on the right-hand side of this formula.
This indicates that $\braket{B_\theta}$ is produced by the effect of $\alpha_{\varphi\varphi}$ against the counter-effects of $\alpha_{\theta\varphi}$ and $\gamma_r$ at the base of the convection zone and at low latitudes.

\section{Discussion}\label{sec:discussion and conclusion}
In Section \ref{sec:mean-field analysis}, the mean-field parameters such as the $a$-tensor (Figure \ref{fig:alpha}) and the turbulent magnetic diffusivity $\beta$ (Figure \ref{fig:beta_comp}) are obtained.
The structure of a large-scale magnetic field is also obtained at a steady polarity period (Figure \ref{fig:average_B}), and its induction (Figures \ref{fig:MS_FA} and \ref{fig:induce}) is analyzed.
These results are used to infer the mechanism which maintains a large-scale field at high Reynolds numbers, which reverses their polarities. 
The cause of the enhancement of $\gamma_r$ in comparison to the previous studies \citep{2011ApJ...735..46P, 2015_ApJ_809_149P,2016_AdSpR_58_1522P} is also discussed.
\subsection{The Large-Scale Magnetic Field in High Magnetic Reynolds Number}\label{subsec:large scale duffuse}
The magnetic energy of the large-scale field decreases from case Low to case Mid and remains at the same level in case High (see Figure \ref{fig:bar}). 
This is explained by using the explicit diffusivities and turbulent magnetic diffusivities (Figure \ref{fig:beta_comp}).

The general effects of the explicit diffusivity and turbulent magnetic diffusivity on large and small-scale magnetic fields are discussed here before discussing the results of this study.
Explicit viscosity and magnetic diffusivity contribute to the smoothening of the spatial and temporal fluctuations of the velocity and magnetic field.
Consequently, turbulence and small-scale dynamo are suppressed by explicit diffusivities, and a large-scale magnetic field is maintained. 
Conversely, the suppression of the turbulence and small-scale dynamo, which distract the large-scale magnetic fields, is less efficient when the explicit diffusivities become smaller.
Consequently, the large-scale magnetic field is distracted.
This implies that the energy of the large-scale magnetic field decreases while that of the small-scale magnetic field increases. 
The turbulent magnetic diffusion is expressed in the induction equation of a large-scale field as:
\begin{eqnarray}
  \frac{\partial \braket{\bm{B}}}{\partial t}=[\cdots]-\nabla\times(\beta\nabla\times\braket{\bm{B}}),
\end{eqnarray} 
whereas it is expressed in the induction equation of the small-scale field with a plus sign, as:
\begin{eqnarray}
  \frac{\partial \bm{B}'}{\partial t}=[\cdots]+\nabla\times(\beta\nabla\times\braket{\bm{B}}).
\end{eqnarray} 
This indicates that the energy dissipated from the large-scale field by the turbulent diffusion, is converted into a small-scale field. 
Note that, not all the energy lost from the large-scale magnetic field due to turbulent diffusion is converted to that of the small-scale magnetic field.
A lesser amount of energy present in the large-scale magnetic field is converted into a small-scale magnetic field when the turbulent magnetic diffusivity decreases.

The simulation results demonstrate a decrease in the energy of the large-scale magnetic field and an increase in that of the small-scale one from case Low to case Mid (Figure \ref{fig:bar}), where the Reynolds number increases between these cases.
The explicit viscosity, $\nu$, and the magnetic diffusivity, $\eta$, between these cases demonstrate larger differences than the turbulent magnetic diffusivity (Figure \ref{fig:beta_comp}). 
This result can be explained by the effect of the explicit viscosity and magnetic diffusivity, as described above.
Case Mid has smaller explicit diffusivities than case Low (Figure \ref{fig:beta_comp}), and the magnetic field of case Mid, thus forms a less large-scale structure. 
Consequently, the energy of the large-scale magnetic field decreases, and that of the small-scale magnetic field increase in case Mid.

However, the energy of the large-scale magnetic field remains at the same level from case Mid to case High, and the energy of the small-scale field increases. 
Only the turbulent magnetic diffusivity, $\beta$ changes between these cases, although there is no change in the explicit diffusivities (Figure \ref{fig:beta_comp}).
This can be attributed to the effect of the turbulent magnetic diffusivity, as explained earlier.
The decrease in the turbulent diffusivity indicates the suppression of energy conversion toward the small-scale field.
The maintenance of the magnetic energy of the large-scale field from case Mid to case High (Figure \ref{fig:bar}) can thus be explained by the decrease in the turbulent diffusivity, which suppresses the energy converted from a large-scale magnetic field.

The decrease in the turbulent magnetic diffusivity in case High, results in the large-scale magnetic field being maintained.
Here we refer to the previous study on the large-scale magnetic field with a high magnetic Reynolds number. 
\citet{2014ApJ...789..70P} calculate kinematic dynamo with high magnetic Reynolds number, which gives 2.5D (two-dimensional and three-components) fluid velocity.
They presented the suppression principle, which states that the construction of a large-scale magnetic field for a higher magnetic Reynolds number requires the suppression of the small-scale dynamo. 
Otherwise, the dynamo scale shifts towards a smaller one for higher magnetic Reynolds numbers, resulting in the destruction of the large-scale fields.
\citet{2014ApJ...789..70P} conducted kinematic calculations, in which the feedback from the magnetic field to the flow field was not considered.
Our results for the magnetic energies of the large-scale and small-scale fields do not correspond to the results of the suppression principle, in that the large-scale magnetic field in case High is maintained without suppressing the small-scale dynamo.
This is because the magnetic field exceeds the equipartition strength with velocity in the small-scale field in case High (Figure \ref{fig:spectra}), and kinematic approximation is no longer applicable.
In this case, the Lorentz force feedback from the small-scale magnetic field to the velocity becomes important.
This effect is thought to be incorporated into the turbulent magnetic diffusivity through the small-scale velocity in the turbulent EMF (Equations (\ref{eq:eq8}) and (\ref{eq:eq23})).
Therefore, our study states that a large-scale magnetic field can be maintained even when the small-scale dynamo works efficiently with high magnetic Reynolds numbers.
\citet{2016Sci...351...1427} reported that an efficient small-scale dynamo suppresses the small-scale flow which destroys the large-scale magnetic field.
Our analysis proposes that the destruction can be incorporated into the turbulent magnetic diffusivity in the mean-field electrodynamics.
The estimation of the turbulent magnetic diffusivities quantitatively supports the effect proposed by \citet{2016Sci...351...1427}.

In our simulation, the turbulent magnetic diffusivity significantly decreases with the number of grid points.
We cannot confirm that this tendency continues in the further large numbers of grid points.
Whether the suppression of turbulent magnetic diffusivity and consequent maintenance of the large-scale magnetic field occur at further high Reynolds numbers are needed to be investigated.

\begin{figure}[!]
  \epsscale{0.9}
            \plotone{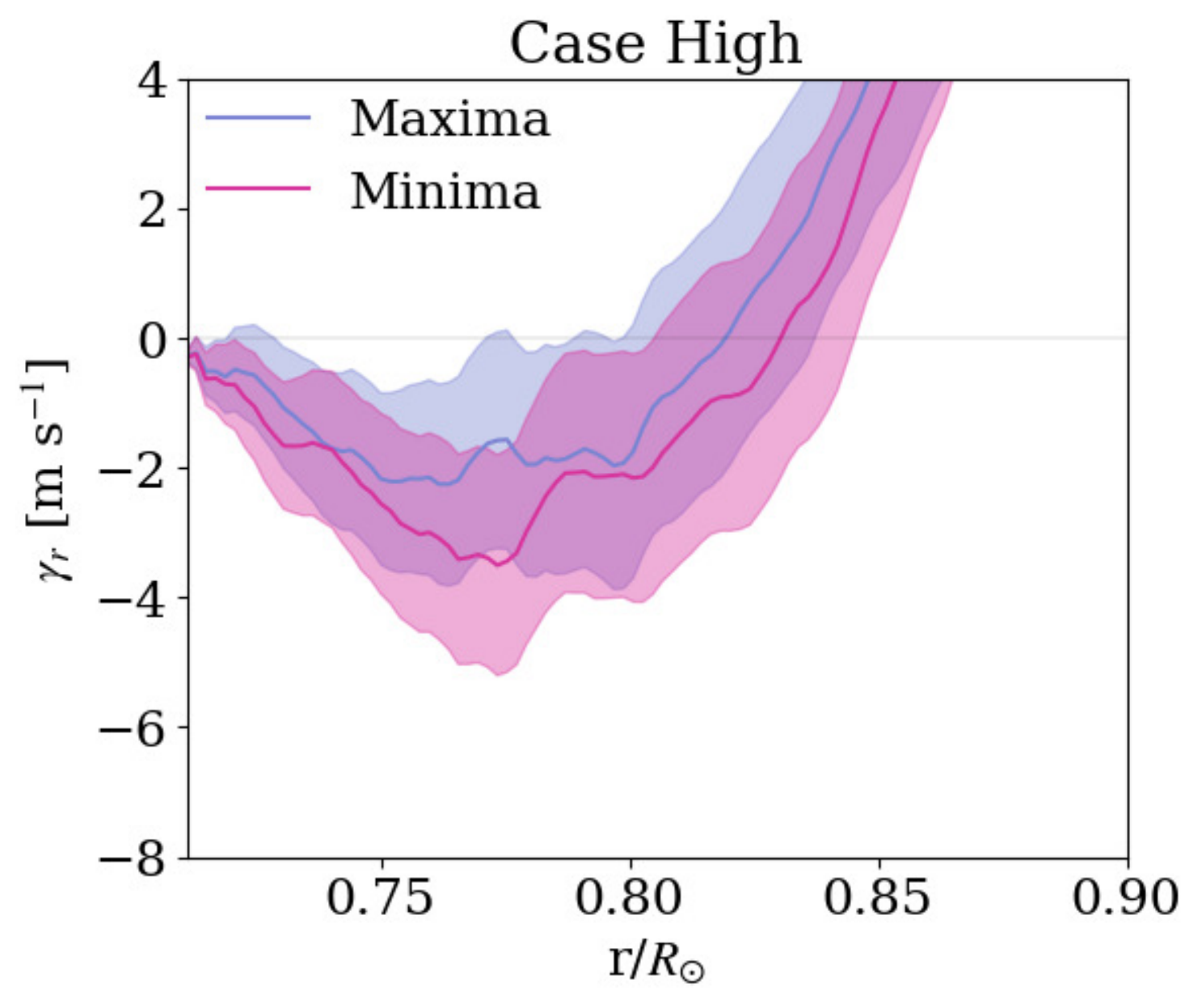}
  \caption{Radial plot of $\gamma_r$ during the maxima and the minima for the case High. The $1\sigma$ section is represented by the shadow. These are taken as latitudinal averages at low latitudes ($0\leqq \Theta \leqq30\arcdeg$). }
  \label{fig:gamma}
\end{figure}

\subsection{Polarity Reversal}\label{subsec:polarity reversal}
All the cases in our simulations show the polarity reversal of the large-scale magnetic field (see Figure \ref{fig:butterfly}).
The large-scale magnetic field comprises the spatial structures presented in Figure \ref{fig:average_B} during the interreversal period. 
Both $\braket{B_\varphi}$ and $\braket{B_\theta}$ are concentrated at the base of the convection zone, where $\braket{B_\theta}$ pointing in the opposite direction, is distributed at the middle depth of the convection zone.
$\braket{B_\varphi}$ is produced by the $\Omega$-effect and fluctuating advection (FA) against the destruction effect from the fluctuating shear (FS), whereas $\braket{B_\theta}$ is produced by the effect of $\alpha_{\varphi\varphi}$ against the effects of $\alpha_{\theta\varphi}$ and $\gamma_r$ at the base of the convection zone and low latitudes (see Subsection \ref{subsec:induction of the mean magnetic field}).

The fluctuating shear reverses the sign of $\braket{B_\varphi}$ (Figure \ref{fig:MS_FA}) based on the obtained relation among the field induction effect.
$\braket{B_\theta}$ which has reversed sign is dominantly induced by $\alpha_{\theta\varphi}$ at the base of the convection zone and at low latitudes. 
This brings the conclusion that $\alpha_{\theta\varphi}$ reverses the sign of $\braket{B_\theta}$.

The $\Omega$-effect can amplify $\braket{B_\varphi}$, and  $\alpha_{\varphi\varphi}$ can amplify $\braket{B_\theta}$ at the base of the convection zone after the signs of both $\braket{B_\varphi}$ and $\braket{B_\theta}$ are reversed by these effects.

Our analysis of the cycle phase dependence of the $\alpha$-tensor and $\gamma$-vector provides additional insights into the triggers of the reversal of $\braket{B_\theta}$.
The $\alpha$-tensor and $\gamma$-vector are estimated using whole periods of the maxima and minima through the procedure described in Subsection \ref{subsec:the alpha tensor} to estimate the cycle phase dependence. 
The definitions of the minima and maxima are presented in Appendix \ref{sec:appendix temporal variation}. 
The \ radial pumping in the magnetic minima is observed to be stronger than the maxima (Figure \ref{fig:gamma}).
$\alpha_{\varphi\varphi}$ and $\alpha_{\theta\varphi}$ which contribute to the induction of $\braket{B_\theta}$ do not demonstrate a clear difference between the maxima and minima (see Appendix \ref{sec:appendix temporal variation}).
Consequently, the strengthened downflow produced by $\gamma_r$ during the minima may trigger the polarity reversal of $\braket{B_\theta}$.

\subsection{The Enhancement of Radial Turbulent Pumping}\label{subsec:gamma}
\begin{figure}[!]
  \epsscale{1}
  \plotone{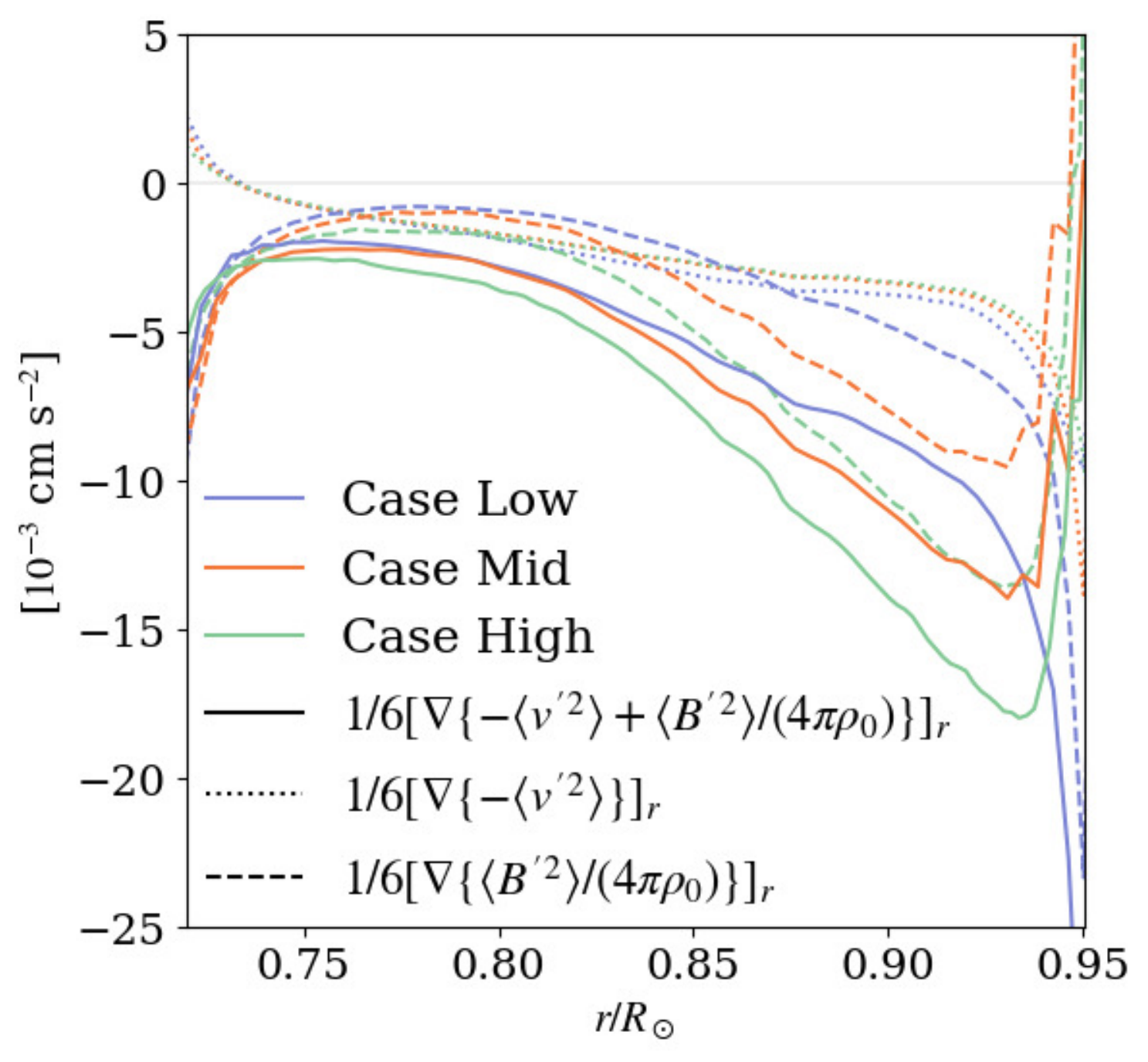}
\caption{Radial plot of the terms which contribute to $\gamma_r$ in Equation (\ref{eq:eqg}). These are averaged over the entire latitudes. The average period is from 5.5 yr to 27.4 yr.}
\label{fig:ana_gamma}
\end{figure}
Our results for the $\alpha$-tensor and $\gamma$-vector differ from those of the previous studies \citep{2011ApJ...735..46P, 2015_ApJ_809_149P,2016_AdSpR_58_1522P} in relative contributions from $\gamma_r$ (Subsection \ref{subsec:the alpha tensor}). 
The contribution of $\gamma_r$ exceeds the components of the $\alpha$-tensor. 
The model of the $\gamma$-vector in terms of the small-scale velocity and magnetic field under the stratification proposed by \citet{radler} as:
\begin{eqnarray}
 \bm{\gamma}=-\frac{1}{6}\nabla\left(\braket{\bm{v}'^2}-\frac{\braket{\bm{B}'^2}}{4\pi\rho_0}\right)\tau_0\label{eq:eqg}
\end{eqnarray} 
is employed to interpret this difference, where $\tau_0$ represents the correlation time.

Figure \ref{fig:ana_gamma} presents each term in Equation (\ref{eq:eqg}).
This indicates that the amplitude of $\gamma_r$ is primarily enhanced by the contribution from the small-scale magnetic field.
The model of the $\alpha$-tensor and $\gamma_\theta$ is also verified, and the results demonstrate that the amplitudes of these terms are mainly determined by the contribution from the small-scale velocity and do not exhibit significant differences between the cases (see Appendix \ref{sec:appendix analytic}).
Our enhancement of $\gamma_r$ when compared to the previous studies \citep{2011ApJ...735..46P, 2015_ApJ_809_149P,2016_AdSpR_58_1522P} can be attributed to the fact that only $\gamma_r$ is affected by the small-scale magnetic field since the small-scale magnetic field is significantly enhanced in our simulation.

\section{Conclusion}\label{subsec:conclusion}
Mean-field analyses are conducted on the results of the global 3D MHD calculations to understand the physics which maintain a coherent large-scale magnetic field at high Reynolds numbers \citep{2016Sci...351...1427}. 
Three cases with increasing Reynolds numbers are considered (Figure \ref{fig:butterfly}). 
The energy of the large-scale magnetic field decreases from case Low to case Mid and remains at the same level in case High (Figure \ref{fig:bar} (a)).

The turbulent magnetic diffusivity in each case is obtained through mean-field analysis \citep{2011ApJ...735..46P,2016_AdSpR_58_1522P}. 
The results (see Figure \ref{fig:beta_3} and \ref{fig:beta_comp}) demonstrate that the turbulent diffusivity monotonically decreases with an increase in the Reynolds numbers.
The maximum value of the turbulent magnetic diffusivity is 7$\times$10$^{11}$ cm$^2$ s$^{-1}$ in case Low and 1$\times$10$^{11}$ cm$^2$ s$^{-1}$ in case High, at the depth, $r\sim0.91R_\sun$. 

The reduction of the turbulent diffusivity in the high-Reynolds number case leads to inefficient transformation since the turbulent diffusivity transforms the magnetic energy of the large-scale field into that of the small-scale field (see Subsection \ref{subsec:large scale duffuse}).
This is considered to be the primary cause for the maintenance of the large-scale magnetic fields at high Reynolds numbers (case High).

The $\alpha$-tensor and turbulent pumping are also obtained through mean-field analysis.
The spatial structures of these components are consistent with those of the previous studies \citep{2011ApJ...735..46P, 2015_ApJ_809_149P,2016_AdSpR_58_1522P} for instance:
(1) The diagonal component of the $\alpha$-tensor is antisymmetric about the equator and has different signs between the surface and the base.
(2) Radial pumping, $\gamma_r$, is downward (negative) in most of the convection zone and upward (positive) in the subsurface. 
(3) The latitudinal pumping, $\gamma_\theta$, is equatorward (positive/negative in the northern/southern hemisphere) in the lower convection zone and polarward in the subsurface.

Our results for the $\alpha$-tensor and $\gamma$-vector differ from those of previous studies \citep{2011ApJ...735..46P, 2015_ApJ_809_149P,2016_AdSpR_58_1522P} in the relative contributions from $\gamma_r$.
The model of the $\alpha$-tensor and $\gamma$-vector under the stratification proposed by \citet{radler} in our simulation indicates that only $\gamma_r$ is significantly affected by the small-scale magnetic field.
Our enhancement of $\gamma_r$ in comparison to that of the previous studies can be attributed to this effect since the small-scale magnetic field is significantly enhanced in our simulation.

$\braket{B_\varphi}$ is concentrated at the base of convection zone ($r/R_{\odot}\la0.8$) and low latitudes ($|\Theta|\la 30\arcdeg$) during the period in which the polarity of the large-scale field is constant. 
$\braket{B_\theta}$ is concentrated at the base of the convection zone ($r/R_{\odot}\la0.8$) and low latitudes ($|\Theta|\la 30\arcdeg$), where $\braket{B_\theta}$ pointing in the opposite direction, is distributed in the middle of the convection zone ($0.8\la r/R_{\odot}\la0.9$) during this period. 
Our estimation of the induction equation demonstrates that the $\Omega$-effect (MS) and advection by fluctuating flow (FA) produce $\braket{B_\varphi}$, and shearing due to the fluctuating flow (FS) works against the production at low latitudes in the base of the convection zone. 
$\braket{B_\theta}$ in this area is produced by shearing due to the fluctuating flow (FS), and advection due to the fluctuating flow (FA) works against it. 
The contributions from the $\alpha$-tensor and turbulent pumping in producing $\braket{B_\theta}$ are also verified.
The results demonstrate that the effect of $\alpha_{\varphi\varphi}$ produces $\braket{B_\theta}$ at low latitudes in the base of the convection zone against the effects of $\alpha_{\theta\varphi}$ and $\gamma_r$.

Based on the obtained relation among the field induction effect, 
the fluctuating shear (FS) reverses the sign of $\braket{B_\varphi}$ (Figure \ref{fig:MS_FA}), and $\alpha_{\theta\varphi}$ plays a significant role in reversing the sign of $\braket{B_\theta}$ in the base of the convection zone (Figure \ref{fig:induce}).
$\Omega$-effect and $\alpha_{\varphi\varphi}$ can amplify both $\braket{B_\varphi}$ and $\braket{B_\theta}$ after the sign of $\braket{B_\varphi}$ and $\braket{B_\theta}$ are reversed.
The cycle phase dependence of $\gamma_r$ (Figure \ref{fig:gamma_r}) indicates that the strengthened downflow due to $\gamma_r$ during the minima may trigger the polarity reversal of $\braket{B_\theta}$.  
This effect may be incorporated into the fluctuating advection (FA) in Equation (\ref{eq:eq28}) at low latitudes ($|\Theta|\la 20\arcdeg$).
We plan to conduct the mean-field dynamo simulation to quantitatively estimate the effect of $\alpha_{\varphi\varphi}$ and $\gamma_r$ on the polarity reversal.

\acknowledgments
H.H. is supported by JSPS KAKENHI Grant No. JP20K14510, JP21H04492, JP21H01124, JP21H04497, and MEXT as a Program for Promoting Research on the Supercomputer Fugaku (Toward a unified view of the universe: from large-scale structures to planets) grant no. 20351188. The results were obtained using the Supercomputer Fugaku provided by the RIKEN Center for Computational Science.
T.Y. is supported by JSPS KAKENHI Grant No. JP15H03640, JP20KK0072, and JP21H01124.

\appendix
\section{Radial distributions of the large-Scale Field}\label{sec:appendix Large-Scale Field}

\begin{figure}[H]
  \epsscale{0.6}
  \plotone{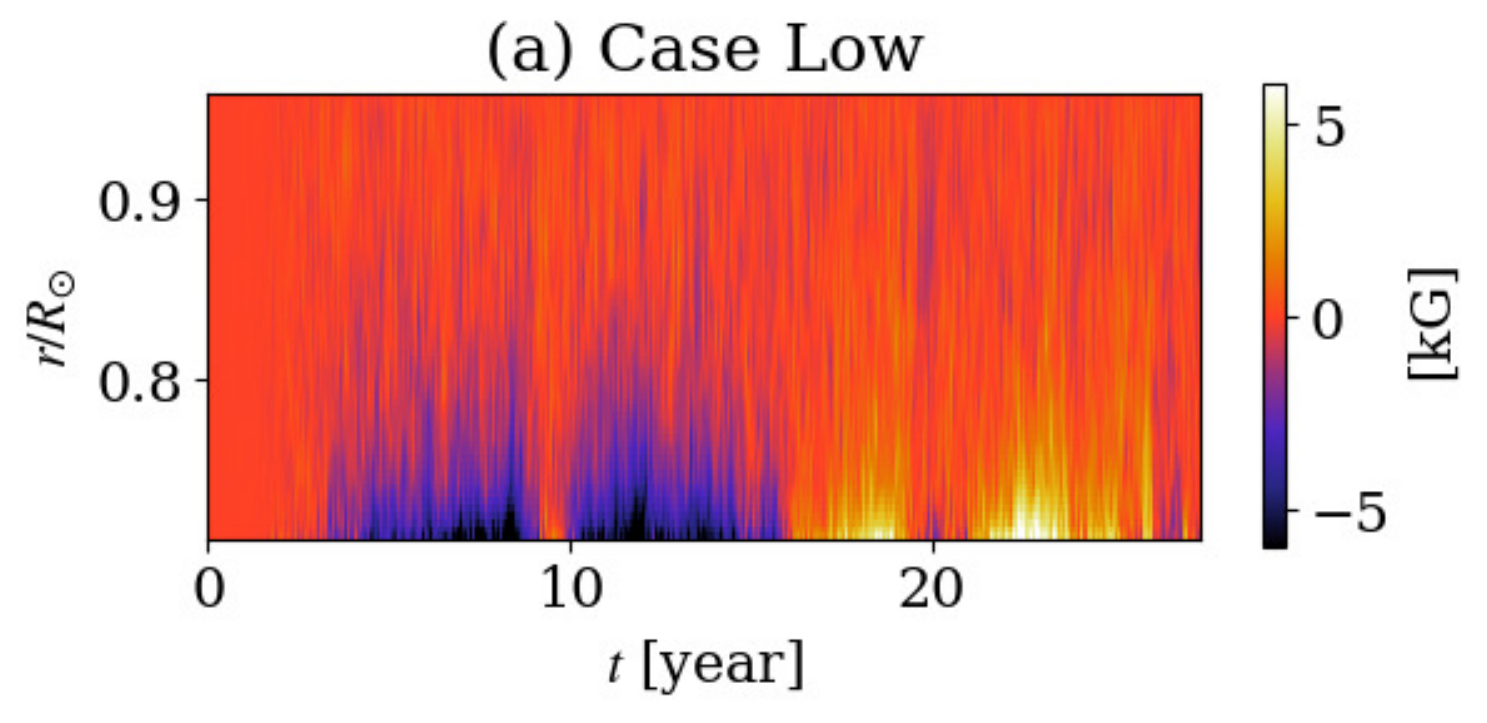}
  \plotone{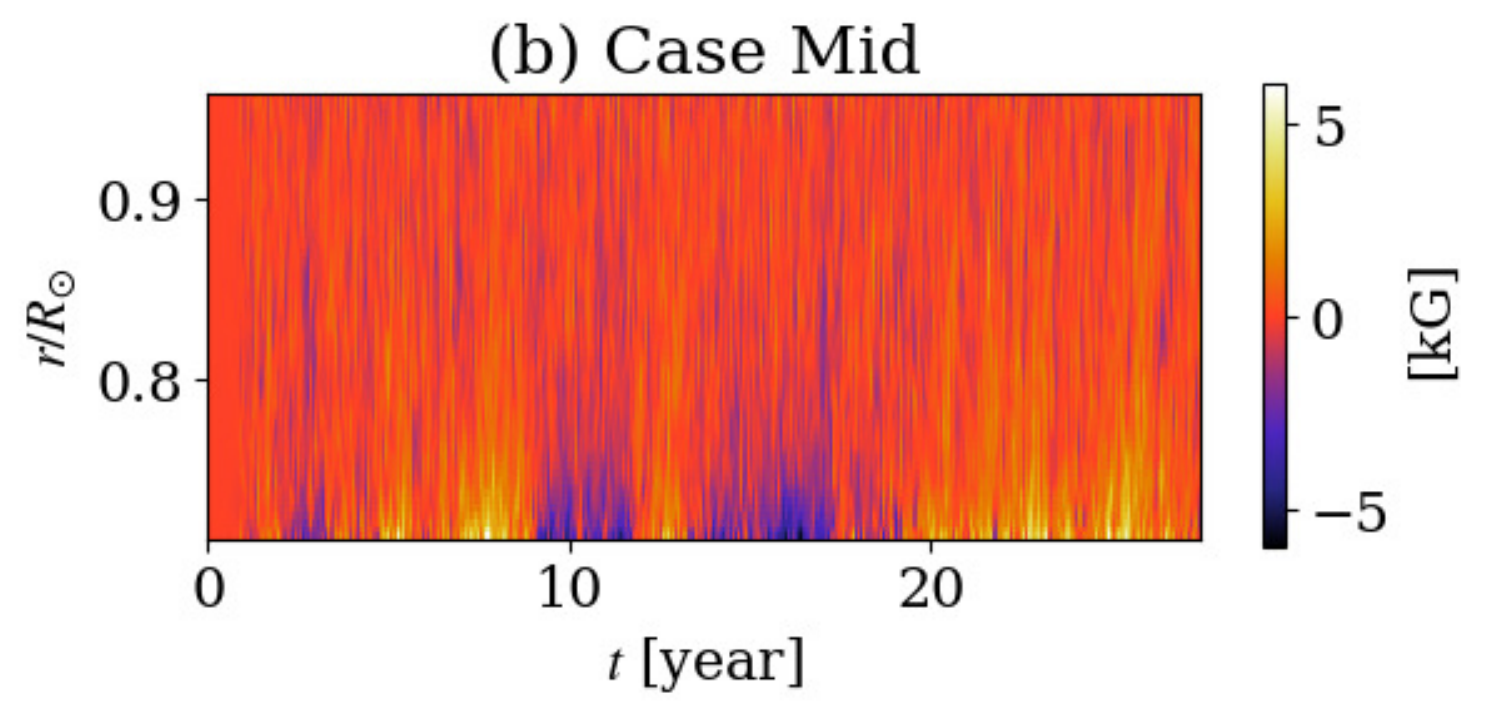}
  \plotone{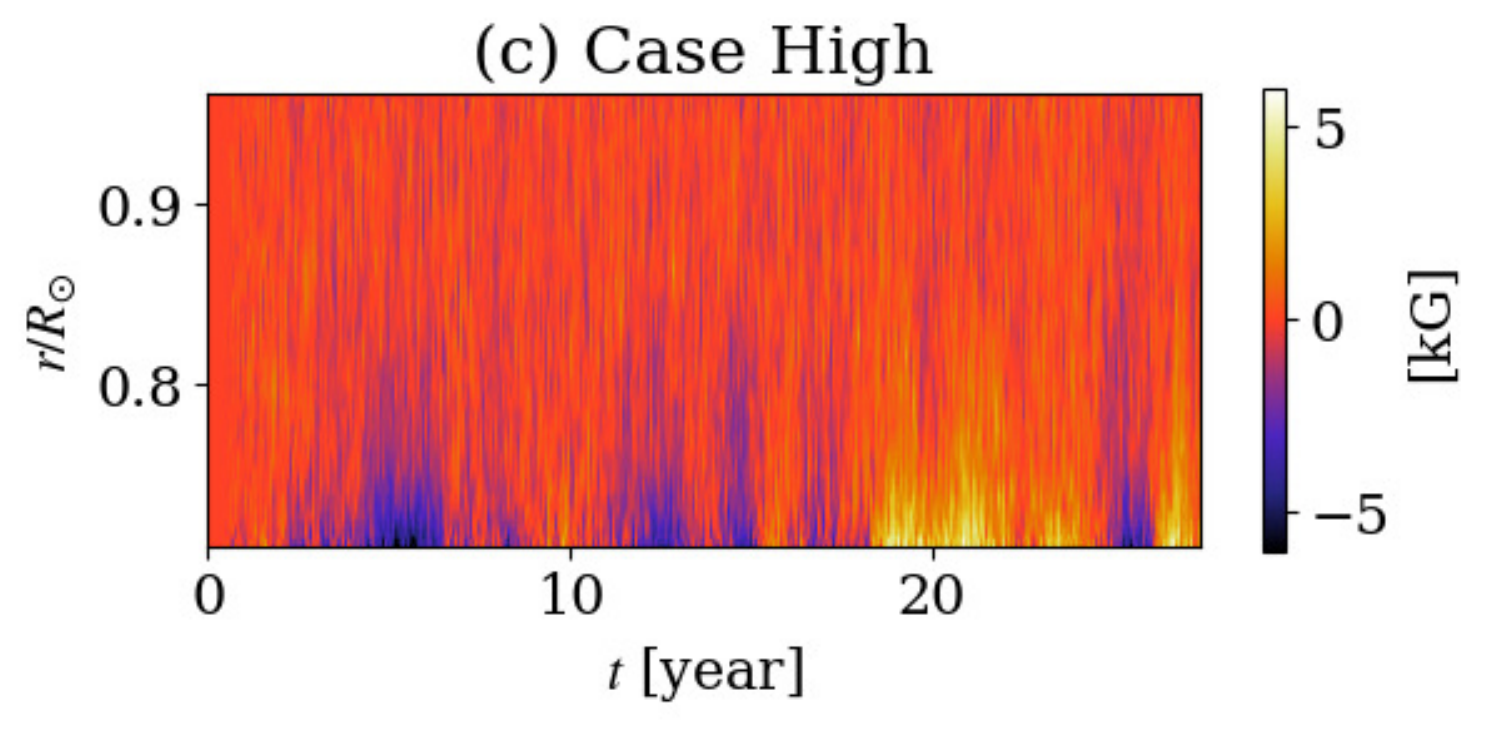}
  \caption{Temporal evolution of the radial distribution of $\braket{B_\varphi}$ at $\Theta=20\arcdeg$.}
  \label{fig:butterfly_2}
\end{figure}

Figure \ref{fig:butterfly_2} depicts the temporal evolution of the radial distribution of the mean toroidal magnetic field, $\braket{B_\varphi}$, at latitude, $\Theta=20\arcdeg$, where the field is concentrated.
The large-scale magnetic field is observed to be concentrated at the base of the convection zone ($r<0.8R_\sun$) in all the cases.

\section{Alpha tensor obtained without using first derivatives of mean magnetic field in the fitting procedure}\label{sec:appendix alpha tensor}

Subsection \ref{subsec:the alpha tensor} presents the result of the $\alpha$-tensor and the $\gamma$-vector (Figure \ref{fig:alpha}) obtained from the fitting method, which includes the first derivatives of the mean magnetic field in the fitting procedure (Equation \ref{eq:eq25}).
The result obtained by the method which does not include the first derivatives of the mean magnetic field in the fitting procedure is presented in this instance, as originally proposed by \citet{2011ApJ...735..46P}.
In this method, the fitting formula for Equation \ref{eq:eq25} is described as follows: 
\begin{eqnarray}
  \mathcal{E}_i(t,r,\theta)=&&\tilde{a}_{ij}(r,\theta)\braket{B_j}(t,r,\theta), \label{eq:ap1}
\end{eqnarray} 
The fitting procedure is almost identical to that in Subsection \ref{subsec:the alpha tensor} except for the designed matrix in SVD \citep[see][for further detail]{2011ApJ...735..46P}.
The $\alpha$-tensor and $\gamma$-vector are automatically determined through Equations (\ref{eq:eq24}) and (\ref{eq:eq26}) after $\tilde{a}$ is estimated by the fitting procedure. 

\begin{figure}[H]
  \epsscale{1.2}
  \plotone{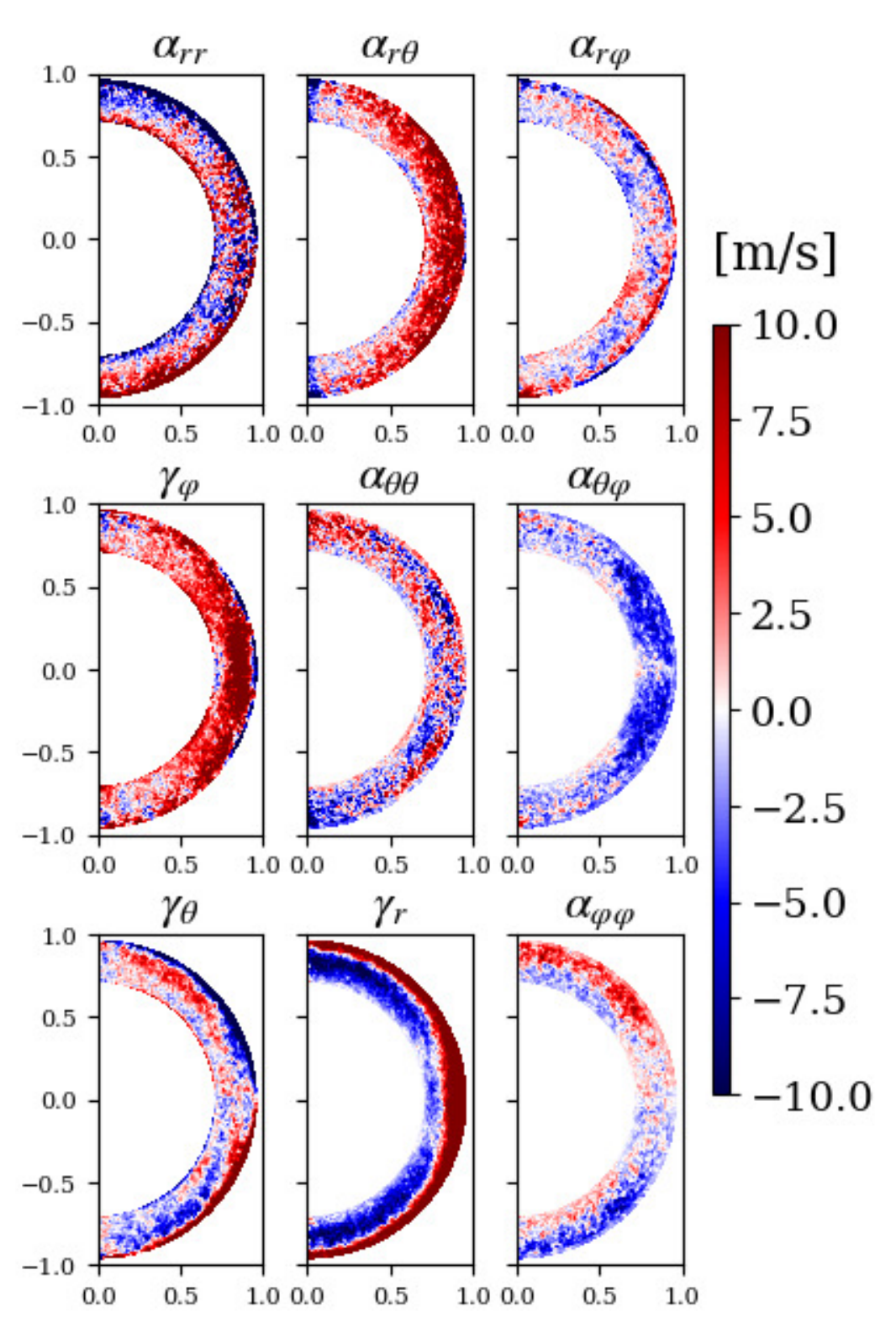}
  \caption{Components of the $a$-tensor extracted from case High plotted in meridional plane. The signs of $\gamma_\theta$, $\alpha_{r\theta}$, and $\alpha_{\theta\varphi}$ differ from those in \citet{2011ApJ...735..46P} owing to the definition of the coordinates. The color bar and scaling are uniform across all the panels.}
  \label{fig:alpha-racine}
\end{figure}

Figure \ref{fig:alpha-racine} presents the obtained $\alpha$-tensor and $\gamma$-vector, and it can be observed that there is no significant difference between Figure \ref{fig:alpha} and Figure \ref{fig:alpha-racine}.

\section{Induction effects}\label{sec:appendix whole terms of the induction equation}

\begin{figure}[H]
  \epsscale{1}
  \plotone{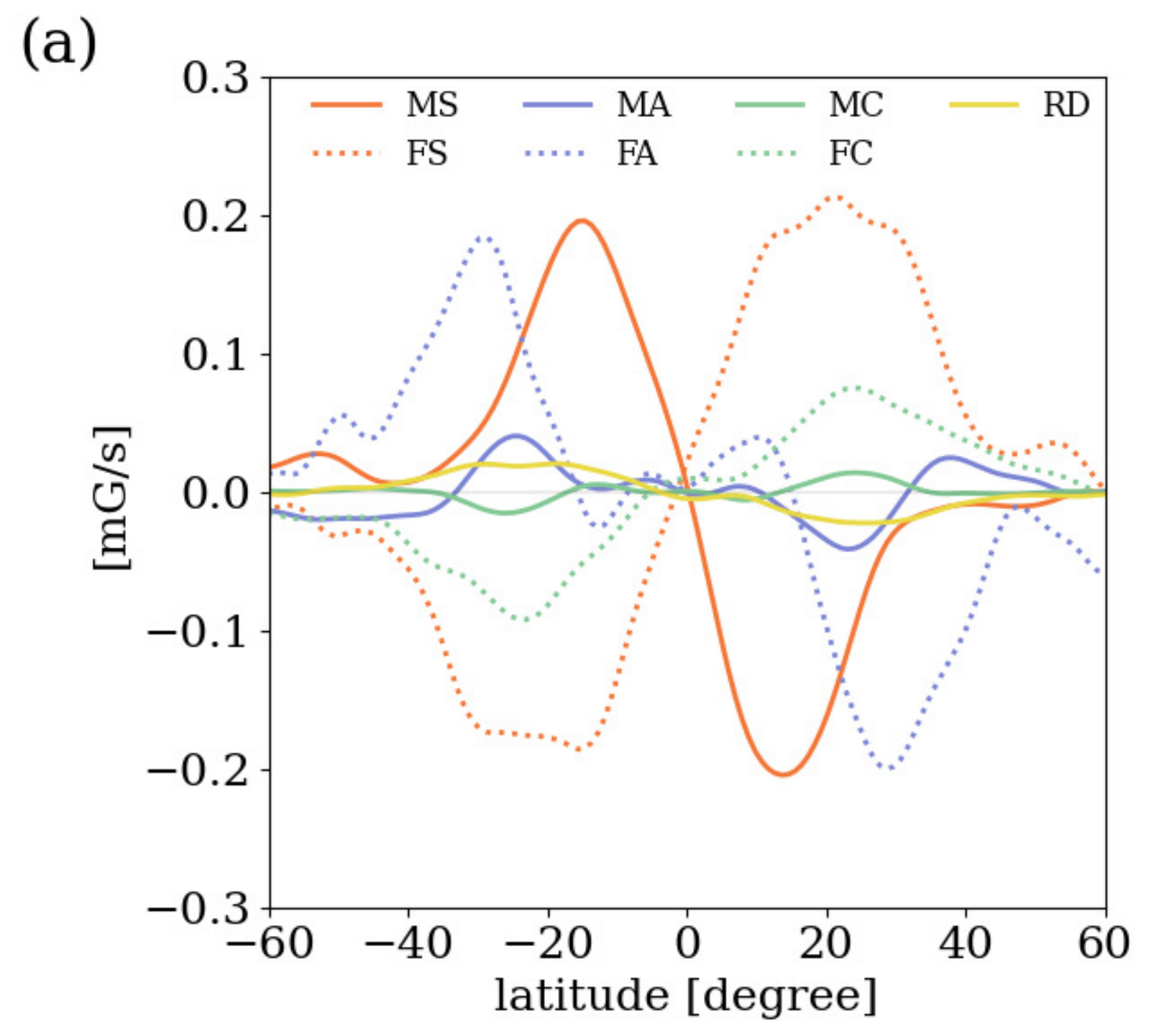}
  \plotone{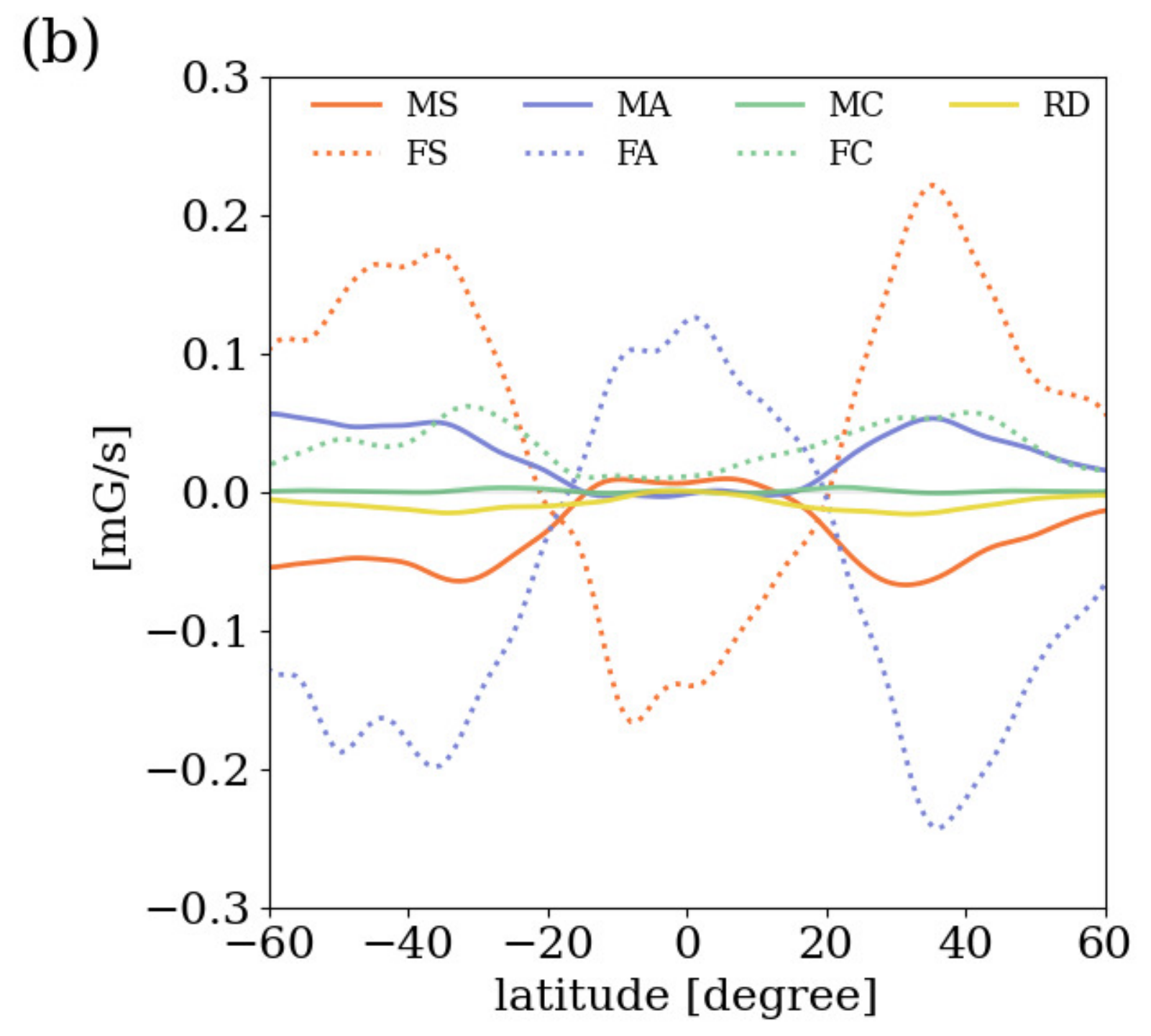}
\caption{(a) Latitudinal plot of time-averaged whole term in Equation (\ref{eq:eq28}) which contribute to the induction of $\braket{B_\varphi}$ at the base of convection zone. ($r/R_{\odot}\leqq0.8$) (b)  Latitudinal cut of whole term in Equation (\ref{eq:eq28}) which contribute to the induction of $\braket{B_\varphi}$ at the base of base of convection zone ($r/R_{\odot}\leqq0.8$). Both (a) and (b) correspond to case Low, which are averaged over $0.71R_\sun\leqq r \leqq 0.80R_\sun$ and filtered by a Gaussian filter with a half-width of $0.01R_\odot$ in the latitudinal direction to read large-scale characteristics.}
\label{fig:whole}
\end{figure}

Figure \ref{fig:whole} presents all the terms on the right-hand side of the induction equation (Equation \ref{eq:eq28}).

\section{Cycle phase dependence of the alpha-tensor and the gamma-vector}\label{sec:appendix temporal variation}

The maxima and minima are defined as the period in which the magnetic energy is larger and smaller than its temporal average over the entire simulation interval. 
The minima includes the reversal phase in this definition.
Only the northern hemisphere is considered for this definition along with the related arguments in Subsection \ref{subsec:polarity reversal} due to the weak parity synchronization in our simulation results.

\begin{figure}[H]
  \epsscale{1}
  \plotone{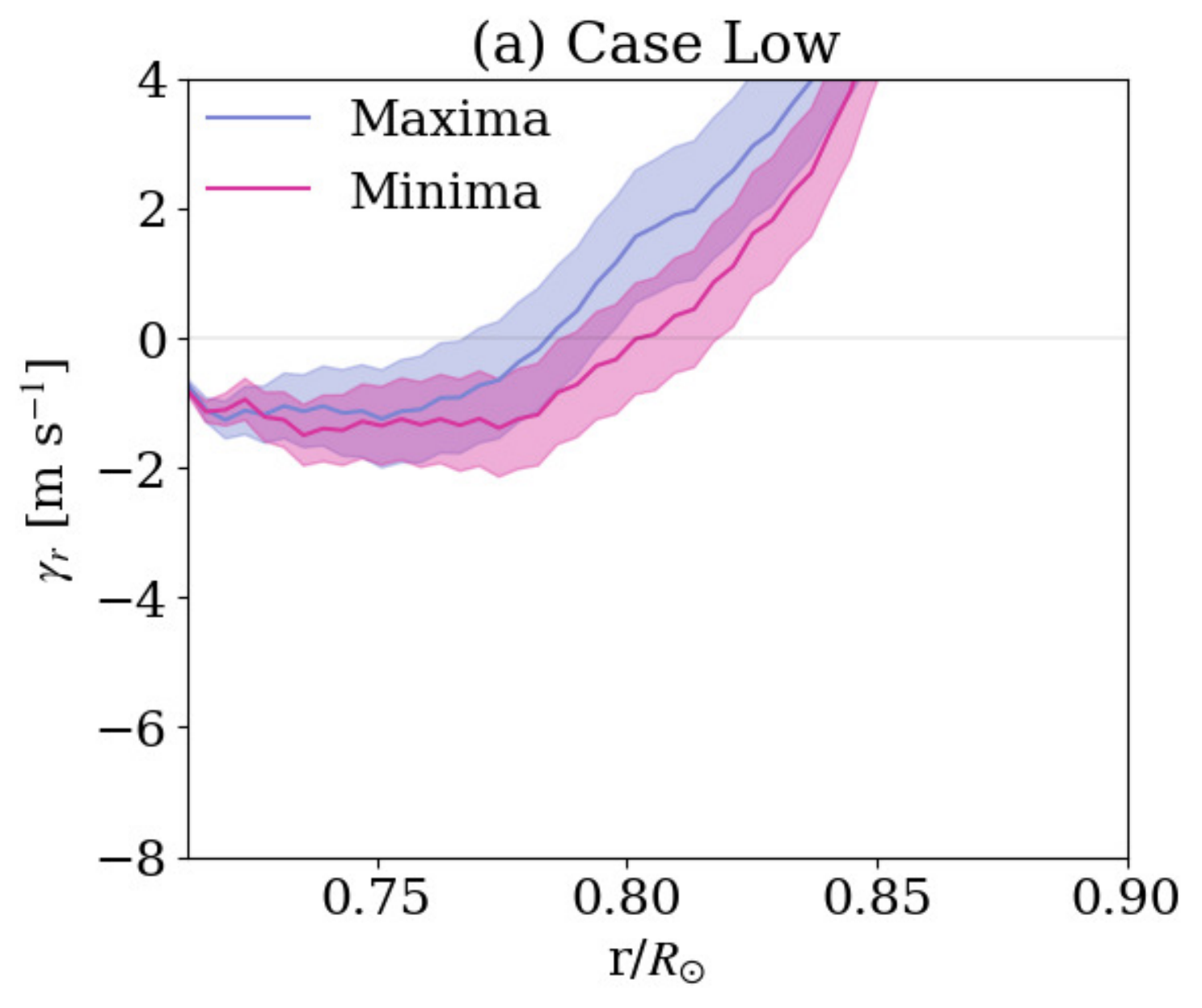}
  \plotone{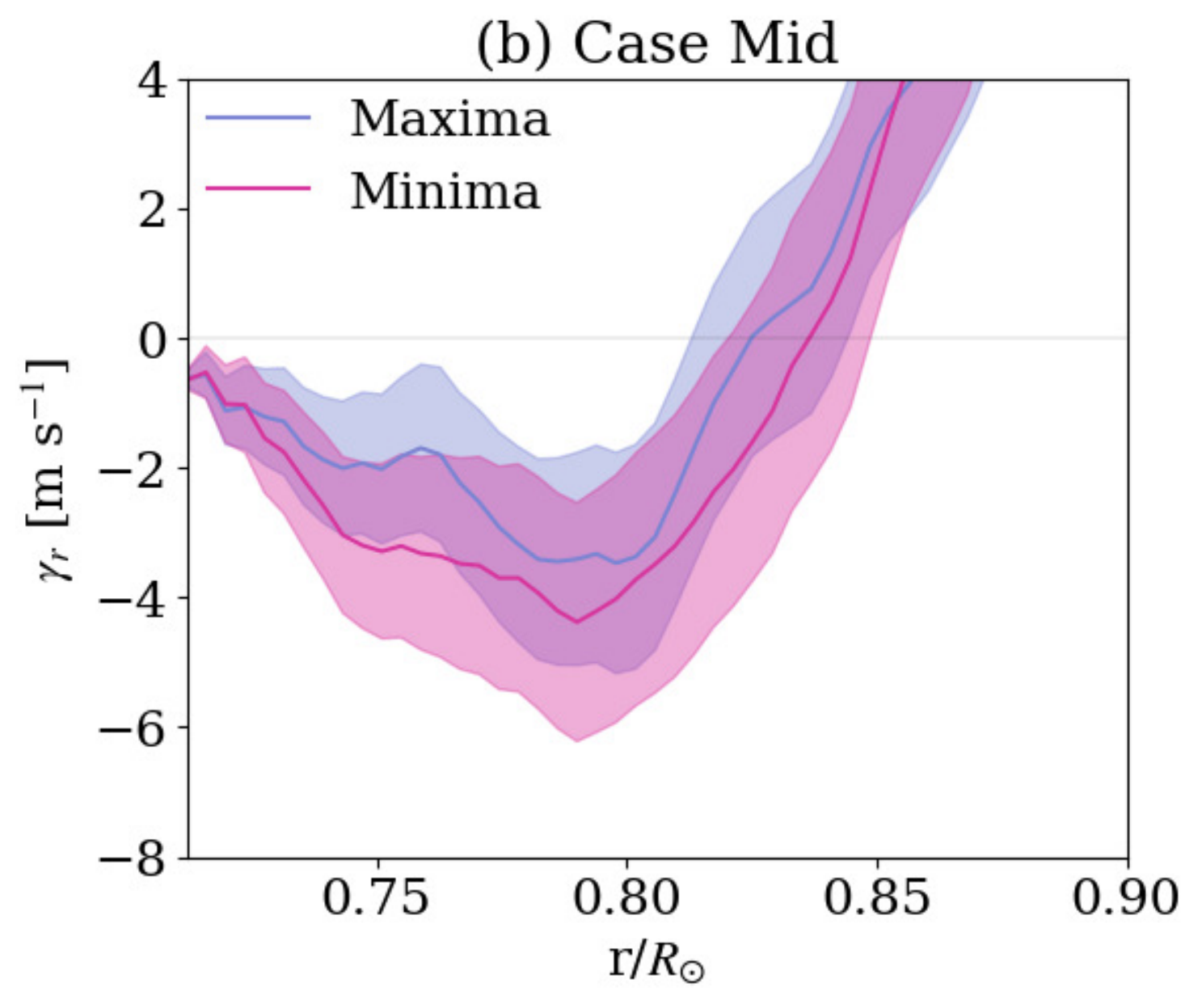}
  \caption{Radial plot of $\gamma_r$ during the maxima and the minima for (a) case Low and (b) case Mid. The $1\sigma$ section is represented by the shadow. Both (a) and (b) show the latitudinal average at low latitudes ($0\leqq \Theta\leqq30\arcdeg$). 
  }
  \label{fig:gamma_r}
  \end{figure}

The radial component of the turbulent pumping $\gamma_r$ during the maxima and minima are shown for the High case in Figure \ref{fig:gamma} (b) and for the other cases in Figure \ref{fig:gamma_r}. 
All the cases exhibit the trend that downward pumping is stronger in the minima than in the maxima by approximately 1$\sigma$ level.

\begin{figure}[H]
  \epsscale{1}
  \plotone{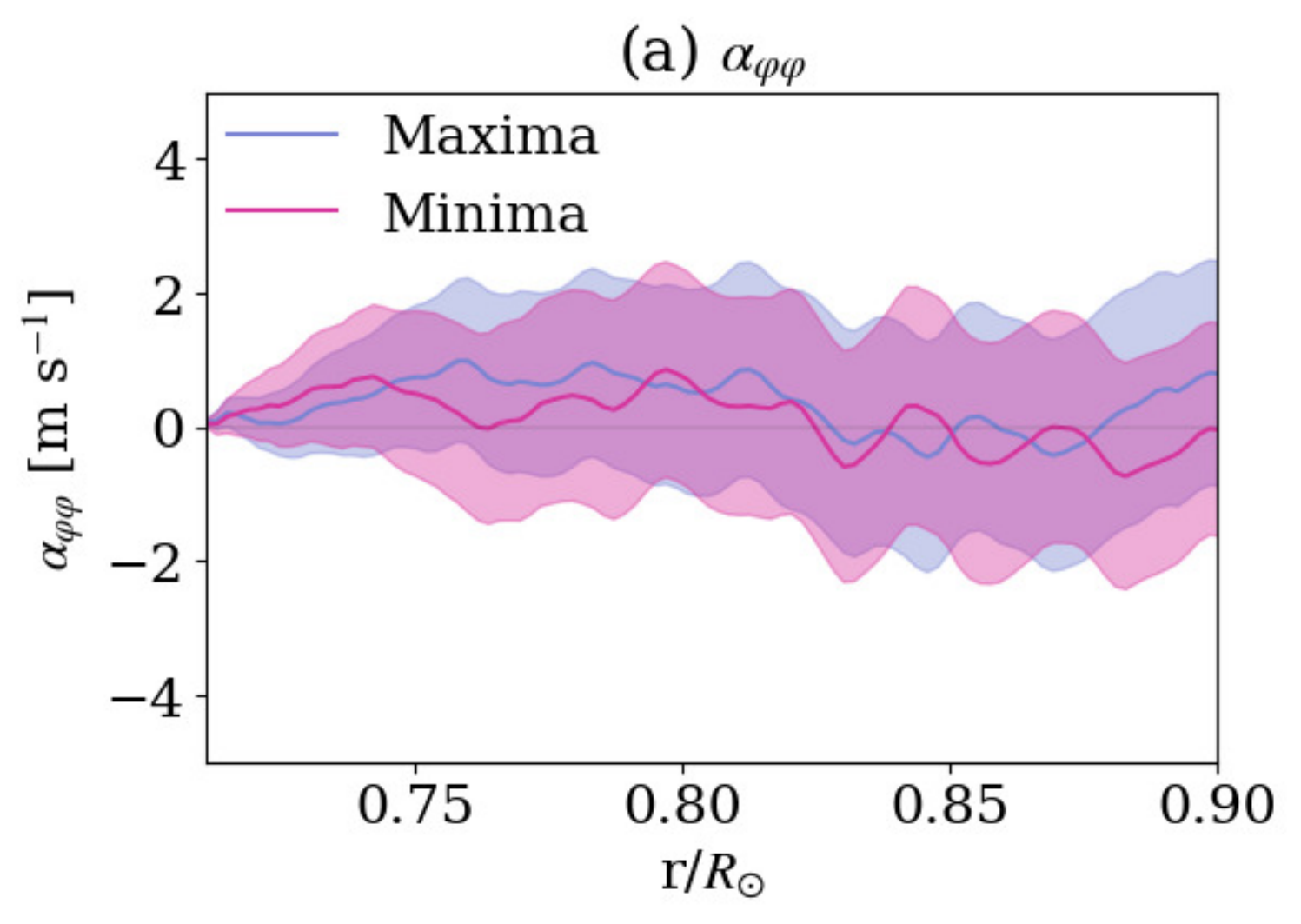}
  \plotone{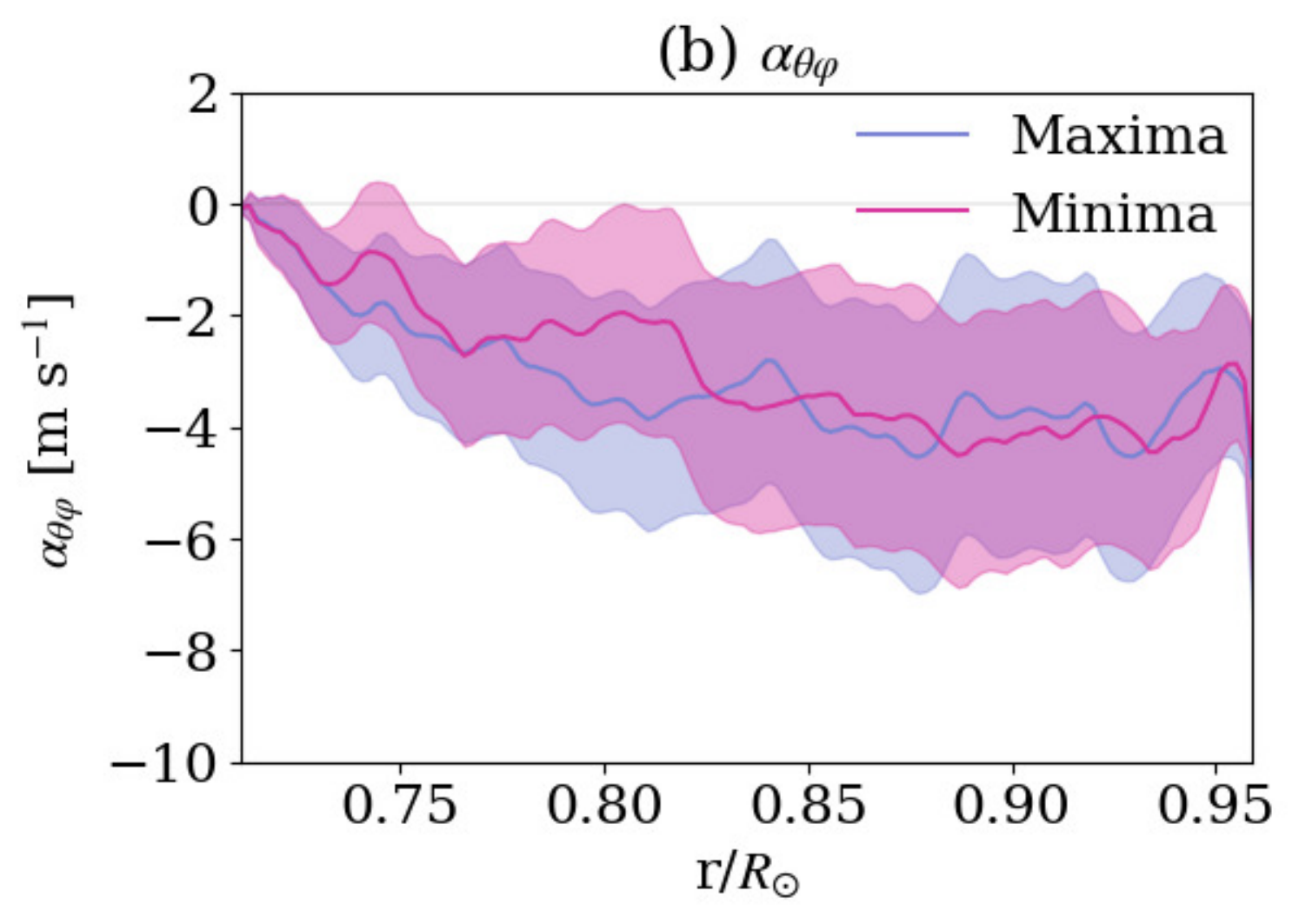}
  \caption{Radial plot of (a) $\alpha_{\varphi\varphi}$ and (b) $\alpha_{\theta\varphi}$ during the maxima and the minima for case High. The $1\sigma$ section is represented by the shadow. (a) shows the latitudinal average at low latitudes ($0\leqq \Theta\leqq20\arcdeg$). (b) shows the latitudinal average at low latitudes ($0\leqq \Theta\leqq30\arcdeg$).}
  \label{fig:alpha_vari}  
\end{figure}

Figure \ref{fig:alpha_vari} shows $\alpha_{\varphi\varphi}$ and $\alpha_{\theta\varphi}$ during the maxima and the minima for the case High. 
Neither $\alpha_{\varphi\varphi}$ and $\alpha_{\theta\varphi}$ do not indicate a clear difference between the maxima and minima.

\section{model of the mean-field coefficients}\label{sec:appendix analytic}

\begin{figure*}[!]
  \epsscale{0.5}
  \gridline{\fig{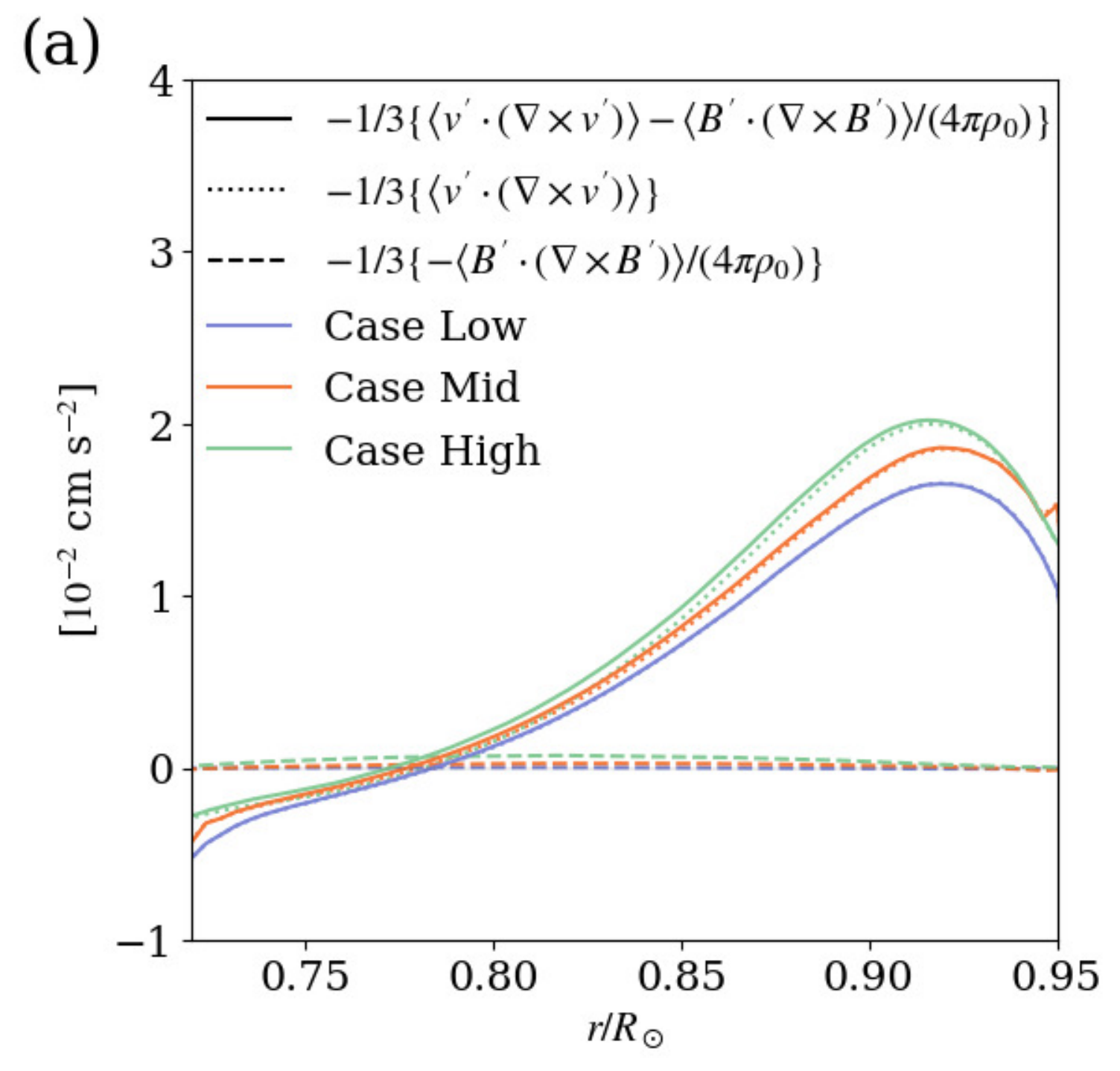}{0.5\textwidth}{}
  \fig{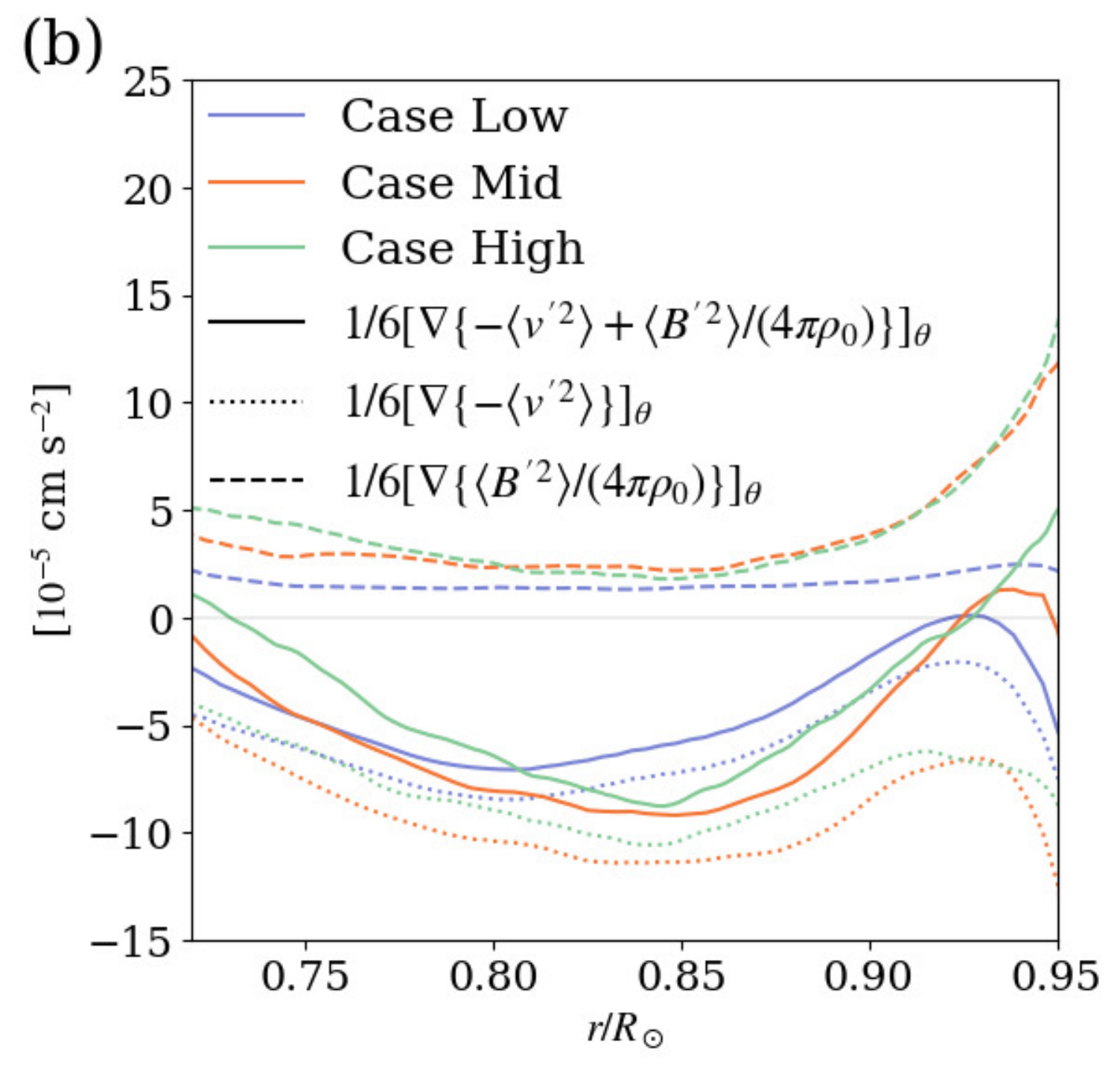}{0.5\textwidth}{}
  }
  \caption{Radial plot of the terms which contribute to (a) $\alpha$ in Equation (\ref{eq:eqA}) and (b) $\gamma_\theta$ in Equation (\ref{eq:eqg}). These are averaged over latitudes between $45\arcdeg<\Theta<90\arcdeg$. The average period is from 5.5 yr to 27.4 yr.}
  \label{fig:A_specific}
\end{figure*}

The models of the $\alpha$-tensor are used in terms of the small-scale velocity and magnetic field proposed by \citet{radler}, in order to interpret the difference between the mean-field coefficients in our study and those of previous works \citep{2011ApJ...735..46P, 2015_ApJ_809_149P,2016_AdSpR_58_1522P},as follows:
\begin{eqnarray}
&\alpha_{ij}=\alpha\delta_{ij},\\  
&\alpha=-\frac{1}{3}\left(\braket{\bm{v}'\cdot(\nabla\times\bm{v}')}-\frac{\braket{\bm{B}'\cdot(\nabla\times\bm{B}')}}{4\pi\rho_0}\right)\tau_0,\label{eq:eqA}
 \end{eqnarray} 
where $\tau_0$ is the correlation time. 
We also use their model for the $\gamma$-vector, that is, Equation (\ref{eq:eqg}).
The results for $\gamma_r$ are presented in Figure \ref{fig:ana_gamma} and show a significant contribution from the small-scale magnetic field (Subsection \ref{subsec:gamma}).
Conversely, the results for $\alpha$ and $\gamma_\theta$ (Figure \ref{fig:A_specific}) exhibit a weak contribution from the small-scale magnetic field.

\newpage
\bibliography{reference}
\bibliographystyle{aasjournal}
\end{document}